\PassOptionsToPackage{table,xcdraw,dvipsnames}{xcolor}
\PassOptionsToPackage{hyphens}{url}
\PassOptionsToPackage{unicode}{hyperref}
\PassOptionsToPackage{naturalnames}{hyperref}

\documentclass[sigconf, screen]{acmart}

\copyrightyear{2024}
\acmYear{2024}
\setcopyright{acmlicensed}\acmConference[HPDC '24]{The 33rd International
Symposium on High-Performance Parallel and Distributed Computing}{June
3--7, 2024}{Pisa, Italy}
\acmBooktitle{The 33rd International Symposium on High-Performance Parallel
and Distributed Computing (HPDC '24), June 3--7, 2024, Pisa, Italy}
\acmDOI{10.1145/3625549.3658682}
\acmISBN{979-8-4007-0413-0/24/06}

\usepackage{enumitem}
\usepackage{multirow} 
\usepackage{tabularx} 
\usepackage{tikz} 

\usepackage[normalem]{ulem} 

\usepackage{pifont}
\newcommand{\cmark}{\ding{51}}%
\newcommand{\xmark}{\ding{55}}%

\usepackage{balance} 

\usepackage[capitalize, noabbrev, nameinlink]{cleveref}
\crefname{section}{\S}{\S}
\Crefname{section}{\S}{\S}

\usetikzlibrary{shapes,arrows}
\usetikzlibrary{decorations.pathreplacing}

\def\custspace{\vspace{3pt}}
\newcommand{\beginbsec}[1]{\custspace\noindent\textbf{#1. \hspace{2pt}}}
\newcommand{\beginbsecnospace}[1]{\noindent\textbf{#1. \hspace{2pt}}}

\newcommand{\CAP}[1]{\scalebox{0.85}{#1}}

\newcommand{\asymb}[1]{\texorpdfstring{$^{#1}$}{}}

\newcommand*\scircled[1]{
\scalebox{0.8}{
\tikz[baseline=(char.base)]{
            \node[shape=circle, text=white, fill=black, draw,inner sep=0.5pt] (char) {#1};}}}

\newcommand{\squishlist}{
 \begin{list}{$\bullet$}
  { \setlength{\itemsep}{2pt}
     \setlength{\parsep}{0pt}
     \setlength{\topsep}{2pt}
     \setlength{\partopsep}{0pt}
     \setlength{\leftmargin}{1em}
     \setlength{\labelwidth}{1em}
     \setlength{\labelsep}{0.5em} } 
}

\newcommand{\squishlistContrib}{ %
 \begin{list}{$\bullet$}
  { \setlength{\itemsep}{2pt}
     \setlength{\parsep}{0pt}
     \setlength{\topsep}{2pt}
     \setlength{\partopsep}{0pt}
     \setlength{\leftmargin}{1em}
     \setlength{\labelwidth}{1em}
     \setlength{\labelsep}{0.5em} }
}
\newcommand{\squishend}{ \end{list}  }

\def\eg{e.g.,~} 
\def\ie{i.e.,~} 

\newcommand{\qt}[1]{``#1''}
\newcommand{\figref}[1]{\cref{#1}}
\newcommand{\secref}[1]{\cref{#1}}

\newcommand{\tabref}[1]{\cref{#1}}
\def\mreqs{M. reqs/s}

\def\dlht{DLHT}

\def\growt{GrowT}

\def\allocator{Allocator}
\def\inlined{Inlined}
\def\hashset{HashSet}

\def\colorhl{\cellcolor[HTML]{C0C0C0}}

\def\colorhl{\cellcolor[HTML]{C0C0C0}} %

\def\gets{Get}
\def\insdel{InsDel}

\newif\ifshowcomment
\showcommentfalse

\ifshowcomment

\newcommand{\todo}[1]{\noindent\textsf{\color{orange}{[{todo: \it #1}]}}}
\newcommand{\boris}[1]{\noindent\textsf{\color{Violet}{[Boris: {\it#1}]}}}
\newcommand{\manos}[1]{\noindent\textsf{\color{purple}{[Manos: {\it#1}]}}}

\newcommand{\pramod}[1]{\noindent\textsf{\color{magenta}{[Pramod: {\it#1}]}}}
\newcommand{\antonis}[1]{\noindent\textsf{\color{OliveGreen}{[Antonis: {\it#1}]}}}

\else

\newcommand{\todo}[1]{}
\newcommand{\antonis}[1]{}
\newcommand{\boris}[1]{}
\newcommand{\pramod}[1]{}
\newcommand{\manos}[1]{}
\fi

\DeclareSymbolFont{extraup}{U}{zavm}{m}{n}
\DeclareMathSymbol{\varheart}{\mathalpha}{extraup}{86}
\DeclareMathSymbol{\vardiamond}{\mathalpha}{extraup}{87}

\newcommand{\bionic}[2][0.5]{%
\StrCut{#2}{ }{\nextword}{\otherwords}%
\exploregroups%
\StrLen{\nextword}[\currlen]%
\edef\halflen{\fpeval{ceil(\currlen*#1)}}%
\bfseries\StrLeft{\nextword}{\halflen}%
\normalfont\StrGobbleLeft{\nextword}{\halflen}\space%
\noexploregroups%
\IfStrEq{\otherwords}{}{}{%
\bionic[#1]{\otherwords}%
}}

\makeatletter
\def\hlinewd#1{%
\noalign{\ifnum0=`}\fi\hrule \@height #1 %
\futurelet\reserved@a\@xhline}
\makeatother
\title[DLHT: A Non-blocking Resizable Hashtable with Fast Deletes and Memory-Awareness]{DLHT: A Non-blocking Resizable Hashtable \\
with Fast Deletes and Memory-awareness}

\acmConference[HPDC '24]{}{June 2024}{Pisa, Italy}

\begin{document}


\author{Antonios Katsarakis\asymb{*}}
\affiliation{%
  \institution{Huawei Research}
}
\author{Vasilis Gavrielatos\asymb{*}}
\affiliation{%
  \institution{Huawei Research}
}
\author{Nikos Ntarmos}
\affiliation{%
  \institution{Huawei Research}
}



\renewcommand{\shortauthors}{A. Katsarakis, V. Gavrielatos, and N. Ntarmos}



\begin{abstract}
This paper presents DLHT, a concurrent in-memory hashtable. 
Despite efforts to optimize hashtables, that go as far as sacrificing core functionality, state-of-the-art designs 
still 
incur multiple memory accesses per request
and block request processing in three cases. First, most hashtables block while waiting for data to be retrieved from memory. Second, open-addressing designs, which represent the current state-of-the-art, either cannot free index slots on deletes or must block all requests to do so. Third, index resizes block every request until all objects are copied to the new index. 
%
Defying folklore wisdom, DLHT forgoes open-addressing and adopts a fully-featured and memory-aware closed-addressing design based on bounded cache-line-chaining. 
This design offers 
\scircled{1} lock-free operations and deletes that free slots instantly,\scircled{2}
completes most requests with a single memory access, 
\scircled{3}
utilizes software prefetching to hide memory latencies, and\scircled{4}
employs a novel non-blocking and parallel resizing. 
%
%
In a commodity server and a memory-resident workload,
DLHT surpasses 1.6B requests per second and
provides 3.5$\times$ (12$\times$) the throughput of the state-of-the-art closed-addressing (open-addressing) resizable hashtable on Gets (Deletes).
%


\end{abstract}

\maketitle



\section{Introduction}
\label{sec:introduction}

\begin{figure}[t]
\vspace{15pt}
\includegraphics[width=0.46\textwidth]{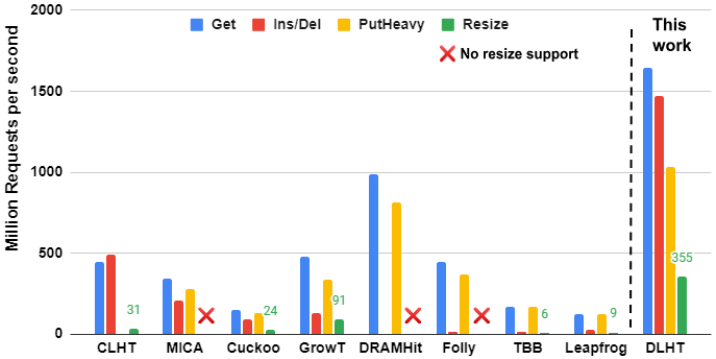}
\caption{Throughput of state-of-the-art hashtables and DLHT with 64 threads in a memory-resident workload (100M objects).}
\vspace{-10pt}
\label{fig:motivation}

\end{figure}

Concurrent in-memory hashtables are essential and versatile data structures in the modern cloud. They are responsible for storing and accessing large amounts of data in main memory via thread-safe \textit{Get}, \textit{Put}, \textit{Insert}, and \textit{Delete} requests.
To ensure requests complete rapidly as the dataset expands, hashtables must be also able to efficiently \textit{Resize} their index.
In-memory hashtables serve a wide spectrum of applications, including in-memory storage, online services, caching, key-value stores, and transactional databases~\cite{Dragojevic:2014, Bronson:2013, DeCandia:2007, Lim:2014, Fan:2013}. 

To meet the ever-growing performance demands~\cite{Elhemali:22, Bronson:2013}, state-of-the-art hashtables from industry and academia offer designs that attain close to a billion requests per second on a single server~\cite{Maier:19, Tudor:15, Li:2016, Li:2017, Lutz:22, Sioulas:19, Schuh:16}. Problematically, their evaluation hints that such high throughput is reachable only under \textit{cache-resident} workloads where accesses are served by hardware caches and seldom reach main memory -- i.e., due to small datasets~\cite{Tudor:15}, data partitioning~\cite{Lutz:22, Schuh:16, Sioulas:19}, or highly skewed accesses~\cite{Maier:19, Li:2016, Li:2017}. 
So we pose the following question:
\textit{Can state-of-the-art in-memory hashtables attain a billion requests per second under a memory-resident workload?}

To answer this question, we evaluate eight state-of-the-art designs over a \textit{memory-resident} workload of 100 million objects accessed uniformly on a commodity server (details in \cref{sec:methodology}).
%
%
%
As shown in \cref{fig:motivation}, almost all hashtables are more than 2$\times$ slower than a billion requests per second. 
%
%
%
The most recent work, DRAMHiT~\cite{Narayanan:23}, is the only one close to the target (in Gets), but its open-addressing design hinders Deletes and Resizes.
Hence, achieving a billion requests per second without forfeiting core functionality on a commodity server remains a challenge for memory-resident workloads.

%
In a deeper inspection (detailed in \cref{tab:core_features} and \cref{sec:background}), 
state-of-the-art designs offer lock-free accesses 
but sacrifice core functionality, incur multiple memory accesses per request, and block processing in three cases.
First, most hashtables stall processing on every request when accessing memory.
%
Second, open-addressing designs offer impaired Deletes that 
either cannot reclaim index slots or 
must cease processing and rebuild the entire index to do so.
%
Third, 
those that support 
index Resizes block every request until all objects are copied to the new index.
%
%
These stalling factors impede the throughput of state-of-the-art hashtables, rendering them \textit{practically blocking} under memory-resident workloads.
In this work, we introduce \dlht, a concurrent hashtable that is \textit{memory access aware} and \textit{practically non-blocking} (i.e., alleviates stalling) to transcend a billion requests per second in memory-resident workloads.
%
Defying folklore wisdom~\cite{Maier:19, Stivala:10, Narayanan:23}, 
\dlht\ forgoes open-addressing and adopts a closed-addressing approach.
Its design is based on bounded cache-line-chaining
and has the following features.
First, it enables lock-free index operations, including deletes with immediate index slot reclamation.
Second, it minimizes memory traffic and completes most requests with a single memory access.
Third, it exploits software prefetching to overlap the memory latency of a request
with productive work on other requests.
%
Finally, it incorporates a novel, non-blocking (but not lock-free) Resize where requests complete with strong consistency while a multi-threaded index migration occurs in parallel.

Unlike state-of-the-art designs that trade core functionality for throughput~\cite{Maier:19, Li:2016, Folly:23, Tudor:15}, DLHT provides a complete set of implemented features to accommodate its clients' needs. Beyond core functionality, this includes 
namespaces, variable-sized key-value pairs, efficient single-thread and hashset variants, as well as pointer APIs that minimize copies.

\begin{table*}[t]
\resizebox{0.95\linewidth}{!}{
\begin{tabular}{l|l|lllll|lll|}
 &
 &
  \multicolumn{5}{c|}{\textbf{Practically non-blocking operations}} &
  \multicolumn{3}{c}{\textbf{Memory access awareness}} \\ \cline{3-10} 
\multirow{-2}{*}{} &
  \multicolumn{1}{c|}{\textbf{
  \begin{tabular}[l]{@{}l@{}}Collision \\ handling\end{tabular}
  }} &
  \multicolumn{1}{c}{\textbf{Gets}} &
  \multicolumn{1}{c}{\textbf{Puts}} &
  \multicolumn{1}{c}{\textbf{Inserts}} &
  \multicolumn{1}{c}{\begin{tabular}[c]{@{}c@{}} \textbf{Deletes}\\ (that free \\ index slots)\end{tabular}} &
  \multicolumn{1}{c|}{\begin{tabular}[c]{@{}c@{}}\textbf{Resizes}\\ (non-blocking:\\ safe Get,..,Del \\ during resize)\end{tabular}} &
  \multicolumn{1}{c}{\begin{tabular}[c]{@{}c@{}} \textbf{Overlaps}\\ \textbf{memory accesses}\\ (i.e., exploits \\ s/w prefetch)\end{tabular}} &
  \multicolumn{1}{c}{\begin{tabular}[c]{@{}c@{}} \textbf{Minimizes} \\ \textbf{memory traffic} \\ (e.g., supports\\ index inlining)\end{tabular}} &
  \multicolumn{1}{c}{\begin{tabular}[c]{@{}c@{}} \textbf{Occupancy} \\ until resize \\ (from \cref{sec:eval:occupancy}\\ with wyhash)\end{tabular}} \\ 
  \hlinewd{3pt}
\multicolumn{1}{l|}{\textbf{GrowT}} &
  \multicolumn{1}{c|}{open-addressing} &
  \multicolumn{1}{c}{{\color[HTML]{036400} \cmark}} &
  \multicolumn{1}{c}{{\color[HTML]{036400} \cmark}} &
  \multicolumn{1}{c}{{\color[HTML]{036400} \cmark}} &
  \multicolumn{1}{c}{{\color[HTML]{9A0000} \begin{tabular}[c]{@{}c@{}}Blocking \\ (until all index \\ is transferred)\end{tabular}}} &
  \multicolumn{1}{c|}{{\color[HTML]{F56B00} \begin{tabular}[c]{@{}c@{}}Parallel,\\ Blocking\end{tabular}}} &
  \multicolumn{1}{c}{{\color[HTML]{9A0000} \xmark}} &
  \multicolumn{1}{c}{{\color[HTML]{036400} \cmark}} &
  \multicolumn{1}{c}{{\color[HTML]{F56B00} \textbf{30-50\%}}} \\ \hline
\multicolumn{1}{l|}{\textbf{Folly}} &
  \multicolumn{1}{c|}{open-addressing} &
  \multicolumn{1}{c}{{\color[HTML]{036400} \cmark}} &
  \multicolumn{1}{c}{{\color[HTML]{036400} \cmark}} &
  \multicolumn{1}{c}{{\color[HTML]{036400} \cmark}} &
  \multicolumn{1}{c}{{\color[HTML]{9A0000} \xmark}} &
  \multicolumn{1}{c|}{{\color[HTML]{9A0000} \xmark}} &
  \multicolumn{1}{c}{{\color[HTML]{9A0000} \xmark}} &
  \multicolumn{1}{c}{{\color[HTML]{036400} \cmark}} &
  \multicolumn{1}{c}{{\color[HTML]{9A0000} \textbf{---}}} \\ \hline
\multicolumn{1}{l|}{\textbf{DRAMHiT}} &
  \multicolumn{1}{c|}{open-addressing} &
  \multicolumn{1}{c}{{\color[HTML]{036400} \cmark}} &
  \multicolumn{2}{c}{{\color[HTML]{036400} \begin{tabular}[c]{@{}c@{}}{\color[HTML]{F56B00}Only upserts}\end{tabular}}} &
  \multicolumn{1}{c}{{\color[HTML]{9A0000} \xmark}} &
  \multicolumn{1}{c|}{{\color[HTML]{9A0000} \xmark}} & 
  \multicolumn{1}{c}{{\color[HTML]{036400} \cmark} {\color[HTML]{F56B00} \begin{tabular}[c]{@{}c@{}} Reorders requests\end{tabular}}} &
  \multicolumn{1}{c}{{\color[HTML]{036400} \cmark}} &
  \multicolumn{1}{c}{{\color[HTML]{9A0000} \textbf{---}}} \\ \hline
  \multicolumn{1}{l|}{\textbf{MICA}} &
  \multicolumn{1}{c|}{closed-addressing} &
  \multicolumn{1}{c}{{\color[HTML]{036400} \cmark}} &
  \multicolumn{2}{c}{{\color[HTML]{F56B00} \begin{tabular}[c]{@{}c@{}}Blocking\end{tabular}}} &
  \multicolumn{1}{c}{{\color[HTML]{F56B00} Blocking}} &
  \multicolumn{1}{c|}{{\color[HTML]{9A0000} \xmark}} &
  \multicolumn{1}{c}{{\color[HTML]{036400} \cmark}} &
  \multicolumn{1}{c}{{\color[HTML]{9A0000} \xmark}} &
  \multicolumn{1}{c}{{\color[HTML]{9A0000} \textbf{---}}} \\ \hline
\multicolumn{1}{l|}{\textbf{CLHT}} &
  \multicolumn{1}{c|}{closed-addressing} &
  \multicolumn{1}{c}{{\color[HTML]{036400} \begin{tabular}[c]{@{}c@{}}{\color[HTML]{F56B00}Unique values}\end{tabular}}} &
  \multicolumn{1}{c}{{\color[HTML]{9A0000} \xmark}} &
  \multicolumn{1}{c}{{\color[HTML]{036400} \cmark}} &
  \multicolumn{1}{c}{{\color[HTML]{036400} \cmark}} &
  \multicolumn{1}{c|}{{\color[HTML]{9A0000} \begin{tabular}[c]{@{}c@{}}Serial, \\ Blocking\end{tabular}}} &
  \multicolumn{1}{c}{{\color[HTML]{9A0000} \xmark}} &
  \multicolumn{1}{c}{{\color[HTML]{036400} \cmark}} &
  \multicolumn{1}{c}{{\color[HTML]{9A0000} \textbf{1-5\%}}} \\ 
  \hlinewd{3pt}
  \multicolumn{1}{l|}{\textbf{DLHT}} &
  \multicolumn{1}{c|}{closed-addressing} &
  \multicolumn{1}{c}{{\color[HTML]{036400} \cmark}} &
  \multicolumn{1}{c}{{\color[HTML]{036400} \cmark}} &
  \multicolumn{1}{c}{{\color[HTML]{036400} \cmark}} &
  \multicolumn{1}{c}{{\color[HTML]{036400} \cmark}} &
  \multicolumn{1}{c|}{{\color[HTML]{036400} \cmark}} &
  \multicolumn{1}{c}{{\color[HTML]{036400} \cmark}} &
  \multicolumn{1}{c}{{\color[HTML]{036400} \cmark}} &
  \multicolumn{1}{c}{{\color[HTML]{036400} \textbf{61-72\%}}} \\ 
\end{tabular}

} 
\vspace{5pt}
\caption{Key features for memory-resident performance across state-of-the-art concurrent in-memory hashtables and \dlht.}
\label{tab:core_features}
\vspace{-10pt}
\end{table*}

We extensively evaluate DLHT on a commodity server using micro-benchmarks, sensitivity studies, application examples, and standard single- and multi-key OLTP benchmarks (YCSB, TATP, and Smallbank). We compare DLHT with eight state-of-the-art concurrent in-memory designs. 
DLHT surpasses 1.6B Get (1.4B Inserts/Deletes, 1B Gets/Puts) requests per second. 
This is more than 3.5$\times$ (3$\times$, 2.7$\times$) the performance of the fastest closed-addressing design and
an order of magnitude faster Deletes than open-addressing designs. Finally, the parallel and non-blocking resize of DLHT allows for a population that is 3.9$\times$ faster than the state-of-the-art.

\noindent In short, the contributions of this work are as follows:
\begin{itemize}[leftmargin=*]
    \item \textbf{We spot core performance and functionality features for memory-resident hashtables} and detail how state-of-the-art concurrent hashtables fall short on those. (\cref{sec:background})

    \vspace{4pt}
    \item \textbf{We introduce \dlht, a closed-addressing hashtable} that is fully-featured and maximizes throughput via a lock-free bounded cacheline-chaining design. \dlht\ minimizes memory traffic per request, leverages software prefetching to mask memory accesses, ensures good occupancy, offers fast Deletes that free index slots instantly, and is equipped with a parallel and non-blocking index resizing. (\cref{sec:design})

    \vspace{4pt}
    \item \textbf{We extensively evaluate the performance of \dlht} on memory-resident workloads over a commodity server.
    \dlht\ provides 1.66B requests/second 
    %
    and 3.5$\times$ (12$\times$) the throughput of the state-of-the-art resizable closed-address\-ing (open-addressing) designs on Gets (Deletes). (\cref{sec:eval})
\end{itemize}

    
\section{Memory-resident Hashtables}
\label{sec:background}






In \cref{tab:core_features}, we summarize two key performance features for hashtables targeting memory-resident workloads:\scircled{1} \emph{memory access awareness} and\scircled{2} \emph{practically non-blocking} operations. 
We break down those features next and subsequently follow on why state-of-the-art falls short on them.

\subsection{Key features for performance} 
\beginbsec{Memory access awareness}
This includes three guidelines.
First, the design must\scircled{1} \emph{overlap accessing memory with useful work}. 
This can be achieved by software prefetching and then switching to doing useful work, while memory is being accessed \cite{Gharachorloo::1992}.
%
We observe that once this is achieved, the bottleneck tends to shift towards memory bandwidth. Therefore the second guideline is to\scircled{2} \emph{minimize the memory traffic per request}. To do so, the design must strive to approach the ideal limit of accessing a single cache-line per request, while avoiding extraneous write-backs to memory (e.g., Gets should not write). This calls for designs where small values are inlined inside the 
index and Gets are lock-free.

Reduced memory traffic can be achieved by
prematurely resizing the index.
To ensure this is not the case, the third rule is that, given a state-of-the-art hash function,
the design must\scircled{3} \emph{maintain high occupancy} -- i.e., a high percentage of occupied to total slots before a resize must be performed. 

\beginbsec{Practically non-blocking}
A Get, Put, Insert or Delete to key $K_A$ is deemed to\scircled{4} \emph{be practically non-blocking} when its algorithm guarantees that it does not impede the progress of any concurrent Get, Put, Insert or Delete to a different key $K_B$. The Resize operation is deemed \emph{non-blocking} when it does not block all other operations until all objects are copied to a new index. The latter is crucial for memory-resident workloads with hashtables that can span gigabytes of data.


    
    

\subsection{Shortcomings of state-of-the-art hashtables} 
From \cref{fig:motivation}, we focus on the five fastest concurrent in-memory hashtables: MICA~\cite{Li:2016, MICA2:19}, CLHT (lock-free)~\cite{Tudor:15}, GrowT~\cite{Maier:19}, Meta's Folly~\cite{Folly:23} and DRAMHiT~\cite{Narayanan:23}. 
In \cref{tab:core_features}, we show how each design falls short on satisfying the performance features, by
handling memory accesses inefficiently, blocking excessively, and even sacrificing core functionality.

\beginbsec{Sacrificed functionality} 
As shown in \cref{tab:core_features}, four out of five fastest hashtables sacrifice Puts, Deletes, or Resizes. Moreover, only MICA supports keys or values beyond 8 bytes (not shown \cref{tab:core_features}).
CLHT assumes that values across different keys in the index are unique.
%
Notably, GrowT, Folly, and DRAMHiT are \emph{open-addressing} hashtables~\cite{Herlihy:2008}. It is well documented that there is no cheap way to perform Deletes in open-addressing~\cite{Knuth:1998, Herlihy:2008, Bender:2021, Peterson:1957, Jimenez:2018, Gao:2005}. All three (GrowT, DRAMHiT, and Folly) perform Deletes through tombstones, which permanently occupy space in the index. These are very fast, but will fill the index after a while. DRAMHiT and Folly do not address that. GrowT moves all live elements to a new index, incurring a large performance penalty (\cref{sec:eval}).


\beginbsec{Memory access awareness}We observe that only MICA and DRAMHiT overlap memory accesses with useful work. Detrimentally, MICA does not inline values in the index, thus mandating multiple memory accesses for a request, even for small values and no collisions. 
DRAMHiT overlaps the memory accesses of a batch of requests provided by a client. However, it performs the requests of the batch out-of-order. Problematically, this can lead to errors in some use cases (e.g., deadlocks in~ \cref{sec:eval:lockm}). Finally, CLHT has low occupancy, as it cannot chain buckets and must Resize after any 4 collisions.

\beginbsec{Practically non-blocking}
MICA adopts a blocking lock-based scheme for updates. 
In contrast, CLHT, GrowT, and Folly support lock-free index modifications. 
DRAMHiT offers lock-free Gets and Upserts, but an application cannot express a pure Put or Insert.
DRAMHiT, GrowT, and Folly struggle with non-blocking Deletes, as previously discussed.
Finally, only CLHT and GrowT can grow on-demand but come with a blocking resize, during which all index operations are stalled. 
\section{Dandelion Hashtable (\dlht)}  \label{sec:design}
This section describes the design of \dlht, which addresses these shortcomings.
First, we present the architecture of \dlht\ (\secref{sec:des:struct}) and its core algorithms 
(\secref{sec:des:alg}). Then we detail how \dlht\
overlaps memory accesses (\secref{sec:des:batch}) and discuss its extra features, including the iterator and namespaces (\secref{sec:des:add}). 

\begin{figure*}[thpb]
    \centering
    \includegraphics[width=1\textwidth]{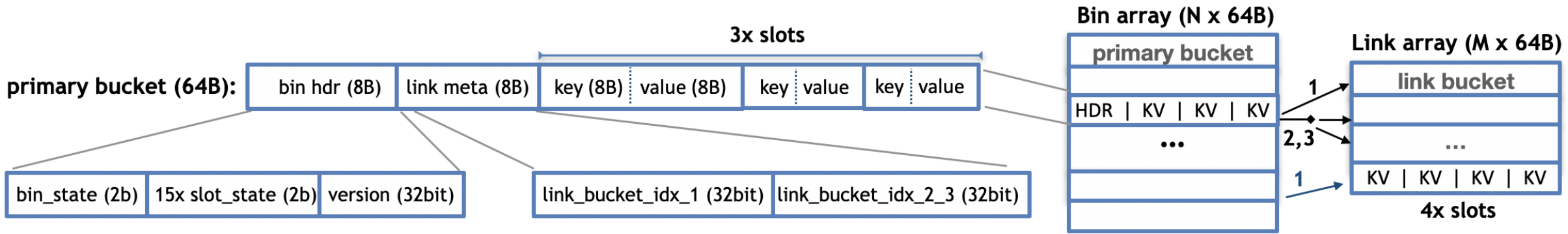}
    \caption{DLHT layout: bins composed of a primary bucket (a bin header and 3 slots) and up to 3 extra linked buckets (4 slots each).}
    \label{fig:dlht}
\end{figure*}


\subsection{Architecture and terminology} \label{sec:des:struct}
In this section, we describe the architecture of \dlht\ (shown in \cref{fig:dlht}) and the basic terms we use throughout the paper. 

\beginbsec{Index} The index is an array of bins.


\beginbsec{Bin} A bin is a chain of buckets. Each chain can have up to four buckets.
Initially, each bin is comprised of a single (primary) bucket. There is an additional array of link buckets, which can be chained to a bin when needed.
There is only a small number of link buckets (by default 8 times fewer than the bins), ensuring that at any time, most bins only have a single bucket -- the primary bucket.
When we run out of link buckets, we trigger a non-blocking resize (\secref{sec:des:alg:resize}).

\beginbsec{Bucket (64 B)}
A primary bucket has an 8-byte bin header followed by 8-byte link metadata and three 16-byte slots for key-value pairs. Link buckets just have four 16-byte key-value slots. Buckets are always cache-line aligned.

\beginbsec{Bin Header (8 B)}
The first 8-bytes of a primary bucket is the bin header, which stores concurrency metadata.
It contains a 32-bit version, a 2-bit bin state, and a 2-bit state for each of the 15 slots -- i.e., for the primary bucket and up to three link buckets per bin.
Fitting that metadata in 8-bytes enables non-blocking operations (e.g., Inserts or Deletes) in the bin via compare-and-swaps (CASes) on the header.
%
In later sections, we detail how these pieces are used.

\beginbsec{Link Meta (8 B)}
The second 8-bytes of a primary bucket contains two 32-bit link-bucket indexes, which are used for chaining up to 3 buckets from the link bucket array to the bin. The first index chains one bucket while the second index chains two (consecutive in link array) buckets to the bin.




\beginbsec{Slot (16 B)} A key-value pair \emph{slot} comprises two consecutive 8-byte segments, one for the key and one for the value.

\custspace\
\dlht\ has three modes of operation. Each mode uses the slot differently and offers a slightly different API. The three configurations are a result of DLHT's real-world use cases.

\beginbsec{1. \inlined} In this mode, key-value pairs are stored in the slot, \ie they are \emph{inlined}. Keys and values must be at most 8 bytes, so that they can both fit in their respective positions inside the slot. 
The \inlined\ configuration is extensively used by \dlht\ clients. Two examples of these use-cases are:\scircled{1} a pointer cache in a query processing engine,\scircled{2} a map between 4- to 8-byte pointers for a database storage engine.

\beginbsec{2. \allocator} In this mode, we allocate additional memory to store a key-value pair. This mode is used when either the key or the value is larger than 8 bytes. The value is always written in the additional memory, and the slot contains a pointer to that memory, instead of the value. If the key is not larger than 8 bytes, it fits in the slot and is not stored externally. When the key is larger than 8 bytes, the slot stores its 8 least significant bytes (for filtering).
In this mode, \dlht\ takes an allocator as input, as in C++ containers. This mode is used as a primary index by a database storage-engine.

\beginbsec{3. \hashset} \dlht\ can be configured as a \emph{\hashset}, which does not store values, only keys. The restriction is that keys must be at most 8 bytes to fit in the slot. The \hashset\ mode is currently used for semi-/anti-joins and by a database lock manager to lock records, where an insertion of a key, locks one or more records, which are released by deleting the key.

Notably, \dlht\ is suitable for more use-cases than those mentioned above. We further discuss some in \cref{sec:design} and \cref{sec:eval}.

\subsection{Algorithms} \label{sec:des:alg}
In this section, we detail the core algorithms of \dlht, which 
are practically non-blocking, minimize memory accesses, and
ensure the strongest consistency (i.e., linearizability~\cite{Herlihy:1990}).
We focus our discussion on the \inlined\ mode, and at the end of each algorithm, we note the changes for the other modes.

\subsubsection{\underline{Get}: lock-free + \textasciitilde~one memory access + ptr API} \label{sec:des:alg:get}
A Get first hashes the key to retrieve a bin. With closed-addressing, there is a one-to-one relation between a key and a bin: if the key is not in that bin, then the key does not exist in the index. The Get does a linear search over all filled slots of the bin. In the worst case, it searches all 15 slots over four buckets. Most commonly, the key hashes in a single-bucket bin; thus, the entire operation only costs a single memory access. When the key is hashed in a bin with more buckets and is not found in the first bucket, a second memory access is required to fetch the second bucket, and so on.

We use the version of the bucket to read a consistent view of the slot. Specifically, we first read the version of the header, then the 2-bit slot state, the slot, and then again the version. This is a common technique to implement lock-free reads, used in seqlocks~\cite{Lameter:2005} and other non-blocking algorithms~\cite{Scott:2013}.

\beginbsec{\allocator} In \allocator\ mode,  the slot stores a pointer instead of the value. 
Additionally, in \allocator\ mode, keys can be larger than 8 bytes; if so, they are stored inside the allocated memory. In this case, we must dereference the pointer to read the key as part of the concurrency algorithm. 

\beginbsec{\allocator\ pointer API}
In \allocator\ mode, we offer a \emph{pointer API}, where a Get returns a pointer to the value and there is no Put. Rather, the Get is used to modify the value. The alternative would have been a \emph{Get/Put API}, where a Get returns the value and Put overwrites it.
We choose the pointer API because of three scenarios:\scircled{1} when the new value depends on the old value\scircled{2} when only a small part of the value needs to be updated and\scircled{3} when the concurrency scheme can be optimized using context-specific information. Puts that blindly lock and overwrite the entire value would incur 
unnecessary copies or index probes in those cases.

\beginbsec{\hashset}Gets return only if a key exists, as it has no values.

\beginbsec{Bounded chaining}
Recall from the previous section that the default number of available link buckets is 8 times smaller than the bins. This ensures that the average number of memory accesses per Get is bounded and remains close to one.

\beginbsec{Summary}
\dlht\ ensures that most Gets require only one memory access and no write-backs. We implement this with
a well-established lock-free algorithm.
In \allocator\ mode, Gets return a pointer to the value. This can be used to efficiently implement custom modification algorithms.

\subsubsection{\underline{Insert}: \scalebox{0.94}{\textit{à la} \CAP{CLHT} + bounded chains + shadow API}} \label{sec:des:alg:ins}
A bin can have up to 15 slots across four chained buckets.  The header of the bin stores a 2-bit state for each slot. This 2-bit state can take the values of Valid, Invalid, and TryInsert. An Insert first checks that the key is not already present, then picks an empty (\ie Invalid) slot to insert, fills the slot, and changes the state of the slot to Valid. 

From a concurrency perspective, 
the challenge is to ensure that there can never be two successful Inserts for the same key given that two Inserts may work on two different slots. 
We solve this using the CLHT lock-free algorithm.
The key idea is to use the bin header as the synchronization point. 
This is possible, despite a bin in \dlht\ having up to four buckets, as all slot states are packed together in the same 8-byte word. 
Specifically, Inserts CAS the bin header to change the state of a single slot. CASing the header ensures the atomicity of the Insert with respect to other operations (Inserts/Deletes) on the same bin.
Notably, because the slot and its state are on separate words, we must perform the Insert in two steps, using an intermediate state called \emph{TryInsert}. 
The Insert algorithm is as follows. 

\beginbsec{Algorithm}\scircled{1} Read the bin header.\scircled{2} 
Run the Get algorithm, if the key already exists return its value along with the corresponding flag.\scircled{3}
Otherwise, find the first slot that is in Invalid state. If no slot can be found, trigger a resize.\scircled{4} 
CAS the header to transition the state of the slot from Invalid to TryInsert.\scalebox{0.8}{\scircled{4.1}} 
If successful, fill the slot.\scalebox{0.8}{\scircled{4.2}} 
If unsuccessful, start over.\scircled{5}
CAS the header again to transition the state of its selected slot from TryInsert to Valid. If unsuccessful we start over from step 1, but we skip steps 3 and 4.

This algorithm ensures that there can never be two successful Inserts for the same key, as only one can succeed in step 5; the other start over and search the bin for the key (step 2), where it will find the successful Insert.
Also, recall that the header has a 32-bit version. The version is incremented with every CAS. This is required by the Get algorithm, but also protects from the notorious ABA bug \cite{Scott:2013}.

\beginbsec{Chaining buckets}After transitioning a slot to TryInsert state (step 4), it is possible that this slot is on a bucket that has not been chained yet. In this case, we first atomically allocate a link bucket (or two consecutive link buckets) through a Fetch-And-Add and then chain it (them) to the bin with a CAS on the link header -- setting the appropriate link-bucket-index. If no link buckets are left, we trigger a resize.


\beginbsec{\allocator} When in this mode, the Insert algorithm allocates memory in step 4.1 (after transitioning the slot state to TryInsert). Note that it is possible for the Insert to fail after the allocation, either because another thread has just inserted the same key, or because the bin ran out of slots. In both cases, we free the allocated memory, before continuing.

\beginbsec{\hashset} The algorithm skips all actions related to values.

\beginbsec{Transactions}Often transactional protocols have two rounds,\scircled{1}lock some objects and\scircled{2} commit (or abort) \cite{Harris:2010}. 
With the pointer API, locking can be implemented by embedding a lock inside each object value. However, this is not possible for Inserts, as the values do not yet exist in \dlht. 
To solve this, the client could implement locking through an additional \hashset.
To avoid this overhead, we offer two additional API calls:
The first is a \emph{shadow Insert} that inserts the key but keeps it hidden from Get/Put/Deletes, essentially locking it. Internally, the only difference of the shadow Insert is that it transitions the inserted key to a \emph{Shadow} state instead of the Valid state (in step 5). 
Later the transaction commits or aborts the shadow Insert using an API call with a corresponding argument, which transitions the slot state to Valid or Invalid.



\beginbsec{Summary}
\dlht\ uses a variant of CLHT's lock-free Inserts.
In the common case, an Insert requires a single memory access, and two CASes on the bin header. The Insert is also responsible for chaining buckets and allocating memory. Finally, we offer optimized Inserts for transactional protocols.

\subsubsection{\underline{Delete}: lock-free + immediate slot reclaim}
\label{sec:des:alg:del}
Unlike open-addressing designs, \dlht\ instantly reclaims slots on Deletes.
To delete, 
we search the bin to locate the matching slot and CAS the header to change the state of the slot from Valid to Invalid. Hence, Inserts can reuse the slot.

\beginbsec{\allocator}
In the \allocator\ mode, \dlht\ stores pointers in the slots instead of values.
Slots are again instantly reused.
To free the pointer after a Delete, we offer an epoch-based GC, for which the client can opt-in.
Our GC remembers the pointers that must be freed. The client periodically performs a call from all threads to advance the epoch. After moving to new epoch, our GC frees the pointers of the previous epoch. 

\beginbsec{\hashset} The Delete is the same as in the \inlined\ mode.

\subsubsection{\underline{Put}: dw-CAS + transfer keys}\label{sec:des:alg:put}
We offer a Put only for the \inlined\ mode. 
We implement Puts using a double-word CAS (dw-CAS) on the slot. Specifically, the client provides a key and the new value to be written. We first execute the Get algorithm to find the corresponding slot. We then use the value that we just read to dw-CAS the entire slot, with the value provided by the client.

It is possible that between reading and dw-CASing, the slot gets deleted and then reused by a subsequent Insert of another key. In this case, the dw-CAS ensures that the Put will either overwrite the entire slot before the Insert, or fail.

Besides concurrent Inserts and Deletes, we must also handle non-blocking Puts racing with an index Resize. It is important that no Put succeeds in the old index after its slot has been copied to the new index. Otherwise, it would be hard to guarantee linearizability.
%
A straightforward solution to this problem, is reducing inlined keys (or values) to 63 bits and requiring the resize algorithm to set this one bit before a transfer occurs. To avoid sacrificing this bit in every slot, we use the idea of \textit{transfer keys} that we discuss in \secref{sec:des:alg:resize}. 


\beginbsec{\allocator\ \& \hashset} 
In \allocator\ there are no Puts. Instead Gets return pointers for updates (discussed in \cref{sec:des:alg:get}). In \hashset\ mode, there are no values and hence no Puts.

\subsubsection{\underline{Resize}: \scalebox{0.95}{parallel growing + concurrent operations}} \label{sec:des:alg:resize}
An index resize is triggered by an Insert when either all slots of a bin are used or there are no more link buckets to chain to the bin. The \emph{resizer}, \ie the thread triggering the resize, executes the entire resize algorithm and then it performs its Insert in the new index. 
Resizing has three steps:\scircled{1} allocate a new index,\scircled{2} transfer all keys to the new index, and\scircled{3} GC the old index. 
Crucially, resizing does not block other threads from executing their requests.
If another Insert also triggers a resize, then it joins the effort of resizing.

\beginbsec{Allocating new index} 
The growth factor varies from 8 when the index is small (e.g., < 4K bins) to 4 for medium-sized indexes (e.g., < 64M bins) to 2 for larger sizes.
The resizer is responsible to allocate the new index.

\beginbsec{Transfer}
The index is partitioned into chunks of 16K bins. This allows threads to collaborate with minimum synchronization. Each collaborating thread (resizer or not) picks a not-yet-transferred chunk and transfers it. 
We detail this collaboration later in this section.
To transfer a chunk, we iterate through all bins, and for every valid 
slot in a bin we:\scircled{1}
read the slot;\scircled{2}
replace the key in the old index with a transfer key; and\scircled{3}
insert the key into the new index.

\beginbsec{Practically non-blocking operations}
Instead of blocking all operations for the duration of the resize, we only block operations at a bin granularity for the duration of a bin transfer.
Recall that each bin has a 2-bit bin state in its header. The bin state can be in NoTransfer (initial state), InTransfer, or DoneTransfer. Before transferring the bin, we CAS the header, setting the bin state to InTransfer. Once the transfer of a bin is done, we set the bin state to DoneTransfer. 

All index operations first check the bin state. If it is InTransfer they wait until it reaches DoneTransfer. If it is DoneTransfer, they perform the operation in the new index. Note that CASing the bin state of the header ensures that all concurrent Inserts and Deletes will either take place before the transfer begins, or will fail and be retried in the new index. A Get also reads the header to complete a slot read and is similarly retried if the bin state is other than NoTransfer.

Puts read the header in the beginning but do not re-read or CAS the header afterwards, only dw-CAS the slot. 
For this reason, the resize algorithm also replaces the key of the slot with a transfer key, before transferring the slot. A transfer key is a key that can never be hashed in the targeted bin.
Our implementation uses one key for odd and another for even bins.
This ensures that either the transfer will see the Put, or the Put will fail, encountering the transfer key. 
In the latter case, the Put is retried and will find the bin in InTransfer or DoneTransfer. It will wait for the transfer to complete if needed and then it will be performed on the new index. 

\beginbsec{Collaboration}When a thread tries to trigger a resize while a resize is in progress, it becomes a \emph{helper}. Helpers first wait for the new index to be allocated. Then they start transferring any available chunks of bins until there are no more chunks to transfer. Finally, they execute their Insert in the new index. 

\beginbsec{GC old index}
Once the transfer is done, the resizer updates the index pointer, ensuring that all new requests use the new index. Now, the resizer needs to
wait for all threads that have read the old index pointer and are still executing a request, to finish their request. 
To do this, we mandate that threads notify each other when finishing a request. We implement this with a per-thread pointer. When a thread enters \dlht\ (e.g., on a Get), we set the pointer to the current index. Just before the thread leaves \dlht, it sets the pointer to null. Once none of the pointers point to the old index, the resizer can GC it. 
The overhead is two atomic store instructions per request (per batch in practice -- \cref{sec:des:batch}).
Crucially, no programmer intervention is required.




\beginbsec{Summary}Resizes only block operations on a single bin at a time. 
Bin transfers are parallel with minimal synchronization.
We GC the old index, without programmer intervention.

\subsection{Overlapping memory latencies with useful work} 
\label{sec:des:batch}

A side-effect of hashing is that accesses to the index do not follow any specific pattern, nor do they exhibit any locality. This renders hardware techniques ineffective: 
the processor will block on every request waiting for main memory~\cite{Gharachorloo::1992}. Software prefetching can be used to tolerate these latencies, either by batching requests or by using coroutines.

\beginbsec{Batching}
\dlht\ offers a function that executes an array of requests (\ie a batch) while respecting the requested order.
Before executing the requests, \dlht\ loops through the array and issues software prefetches for the bin of each request. This overlaps the memory latencies of all requests in the batch. The technique is inspired by MICA~\cite{Lim:2014}. 
Unlike MICA, our pointer-based API also allows us to prefetch the externally stored values in Allocator mode.


In some software stacks, batching occurs naturally. 
For instance, one of our clients implements distributed transactions over \dlht, where 
each network packet contains multiple requests for different keys.
Instead of looping through the network packet issuing requests one-by-one, the client packages them in a batch, reaping the benefits of \dlht\ batching.
Note that it may be crucial to respect the order of requests in this and similar scenarios,
as locks might need to be grabbed in order to avoid deadlocks (e.g., as in \cref{sec:eval:lockm}).

\dlht\ accepts different request types (i.e., Get, Put, Insert, Delete) on the same batch, respects their order, and amortizes the cost of the index GC (two atomic stores) across the batch.
Moreover, \dlht\ offers the option to 
terminate the execution of a batch on an operation that does not complete successfully.
For example, if an application-level lock cannot be grabbed because a key does not exist, \dlht\ will not perform any subsequent requests of the batch.


\beginbsec{Coroutines}
Some software stacks make the use of batching challenging.
In this case, coroutines can be used alternatively
to mask memory latencies~\cite{He:2020}.
To interoperate with coroutines, we offer a function that prefetches the bin of a key.
The client calls this function prior issuing a request to \dlht, placing a yield in between.  
Consequently,
the client can do useful work, while the bin is fetched from memory.



\subsection{Additional features} \label{sec:des:add}

\dlht\ has been in use for several months. During this time, we have accommodated numerous new features. We detailed some in the previous section, \eg  \dlht's three modes of operation (\secref{sec:des:struct}).
We briefly discuss others next. 
When a feature incurs a performance penalty, the client must explicitly opt-in, such that clients only pay for the features they need.

\subsubsection{Variable size keys and values (in a single index)} \label{sec:des:add:var}

A client can configure \dlht\ to hold key-values of any size.
This allows having one \dlht\ instance for different object types and facilitates batching.
For instance, the client may insert a 2-byte key with a 5-byte value and a 128-byte key with a 1024-byte value.
This feature is available in \allocator\ mode -- the only mode where keys can be larger than 8B. When this feature is enabled, we store the key and value sizes in the allocated memory of each key-value pair. 

This poses a performance challenge. 
%
%
%
For inlined keys (i.e., at most 8 bytes), dereferencing the pointer would double the cost of a Get. 
Recall that each slot, has 8 bytes for the key and 8 bytes for the value. But in the \allocator\ mode, we store pointers instead of a value. As pointers only use 48 bits, we safely overload the 16 most significant bits (MSBs).
We use the four MSBs for the key size (the rest are used in \secref{sec:des:add:name}). Four bits suffice, as keys larger than 8 bytes anyway need to dereference the pointer, which stores the key size.

\subsubsection{Namespaces}\label{sec:des:add:name}
When using a single \dlht\ instance to accommodate different types of keys (e.g., keys from different database tables), it is possible that keys have name conflicts.  
To address this, we introduced \emph{namespaces}.
Specifically, clients can tag a key with a namespace-id, which is an integer that can range from 0 to 4Ki. Keys with different namespace-ids do not conflict inside \dlht.
We allow for 4Ki different namespaces, as we overload the 12 remaining most significant bits (MSBs) from the pointer inside the slot (as discussed in  \secref{sec:des:add:var}).

\subsubsection{Hash functions}
The default hashing algorithm is a simple modulo operation:
\begin{equation*}
    bin\_id\: =\: key\: \%\: number\_of\_bins
\end{equation*}
\dlht\ can also be configured to use wyhash\cite{wyhash:23}. 
We benchmarked many of the most prominent hash functions, including CityHash, xxHash, Murmur3, and FNV1.
We found that wyhash provides the most favorable trade-off between performance and randomness. We chose not to include these benchmarks in the paper, as they are orthogonal to \dlht.

\subsubsection{Iterator: weak and strong snapshots}

We offer an iterator API that allows the client to iterate through all of \dlht's key-value pairs. 
The iterator tracks its current index position.
A Next() function fetches the next pair until it has iterated through all bins.
We provide a strongly-consistent snapshot for the iterator
via an index migration (i.e., a "resize" to a same-size index) that temporarily stalls updates -- until transitioning all bins to a Snapshot state. However, our clients prefer a weakly-consistent snapshot that is non-blocking and does not need a migration.


\subsubsection{Single-thread: synchronization-overhead-free}\label{sec:des:add:single}
A client can opt-in to use \dlht\ with only a single-thread.
\dlht\ includes three sources of overhead for thread-safety:\scircled{1}
lock-free algorithms,\scircled{2}
checking for concurrent resizes, and\scircled{3}
notifying other threads when entering/leaving \dlht. When configured for a single thread, we completely remove the second and third overheads. To alleviate the first overhead, we convert every CAS into a regular store, and downgrade atomic loads/stores into regular loads/stores. We also tried replacing the lock-free algorithms with simplified single-threaded algorithms, but the gain was negligible, and thus we abandoned that avenue.

\subsubsection{Summary}
At the request of its clients, \dlht\ offers variable-size keys/values, namespaces, a sophisticated hashing function, an iterator API, and single-threaded optimizations. The features that incur a performance penalty are disabled by default.

\section{Experimental Methodology}
\label{sec:methodology}

\begin{table}[t]
\resizebox{1.63\linewidth}{!}{
\begin{tabularx}{2\linewidth}{X}
\begin{tabular}{|l|l|}
\hline
Commodity Server & \begin{tabular}[c]{@{}l@{}}Two-socket 18-core Intel Xeon Gold 6254 \\72 h/w threads in total (incl. hyper-threads) \end{tabular}  \\ \hline
Hardware Caches & \begin{tabular}[c]{@{}l@{}} 36$\times$ 1MB L2 | 2$\times$ 24.8MB L3 \end{tabular}  \\ \hline
System Memory & 8$\times$ 32GB DDR4-2933 (256GB in total) \\ \hline
OS / Kernel & \begin{tabular}[c]{@{}l@{}} Ubuntu 20.04.3 | Linux 5.4.0-90-generic \end{tabular}  \\ 
\hlinewd{3pt}
\begin{tabular}[c]{@{}l@{}}  (Source Code of) \\ Evaluated Baselines \end{tabular} &
\begin{tabular}[c]{@{}l@{}} CLHT~\cite{CLHT:18}, MICA~\cite{MICA2:19}, GrowT~\cite{Growt:23}, Leapfrog~\cite{Leapfrog:23},\\ DRAMHiT~\cite{DRAMHiT:23}, Folly~\cite{Folly:23}, Cuckoo~\cite{Cuckoo:13}, TBB~\cite{TBB:23} \end{tabular}  \\  \hlinewd{3pt}
Software Threads & 1, 2, 4, 8, \textbf{16}, 32, 64, 71 \\ \hline
Key Size, Value Size            & \textbf{Fixed}/Variable | \textbf{8B}, 16B, ... , 1.5KB \\ \hline
Hash Function                         & \textbf{modulo}, wyhash \\ \hline
Allocator                    & \textbf{mimalloc (2MB pages)}, malloc  \\ \hline
Number of Bins                 & 16K, .., \textbf{67M} (index size = 4GB), .., 1B (64GB) \\ \hline
Number of Keys                 & 16K, ... , \textbf{100M}, ... , 1B, 1.6B \\ \hline
Resizing                       & \textbf{disabled}, enabled     \\ \hline
Access Patterns                & \textbf{uniform}, skewed     \\ \hline
DLHT modes                     & \textbf{Inlined}, Allocator, HashSet \\ \hline
Batching                       & disabled, batch-size = 1, 2, ... , \textbf{32}, 64, 128  \\ \hline
\begin{tabular}[c]{@{}l@{}} 
Workloads/ \\
Application scenarios/ \\
Benchmarks
\end{tabular}
& \begin{tabular}[c]{@{}l@{}}  \textbf{1. \gets,  2. \insdel},  3. PutHeavy, 4. Resizing, 
\\ 5. Single-thread, 6. Lock manager, 7. CXL emul.
\\ 8. YCSB, 9. TATP and Smallbank, 10. Hash Join 
\end{tabular}
\\ \hline

\end{tabular}
\end{tabularx}
}
\vspace{5pt}
\caption{Experimental configuration (default values in \textbf{bold})}
\label{tab:config}
\vspace{-20pt}
\end{table}


\tabref{tab:config} summarizes the testbed, the 
evaluated hashtables, and the 
variables we study in our evaluation.
Unless stated otherwise, the experiments use the default values (\textbf{bold} in \tabref{tab:config}).
We next 
clarify aspects that do not fit in the table.

\beginbsec{Testbed}
We conduct the experiments on a server with two 18-core CPUs and two threads per core (72 hardware threads in total). We spare a thread for the OS; thus, the hashtables use up to 71 threads.
Across all experiments and for all hashtables, we pin threads as follows. When running fewer than 18 software threads, we pin them all in the first socket. When running between 19 and 36, we balance them among the two sockets. When running more than 36, we use hyper-threads and again balance them among the sockets.

\begin{figure*}[t]
\setkeys{Gin}{width=1.05\linewidth}
\begin{tabularx}{\linewidth}{XXX}


\includegraphics[width=0.34\textwidth]{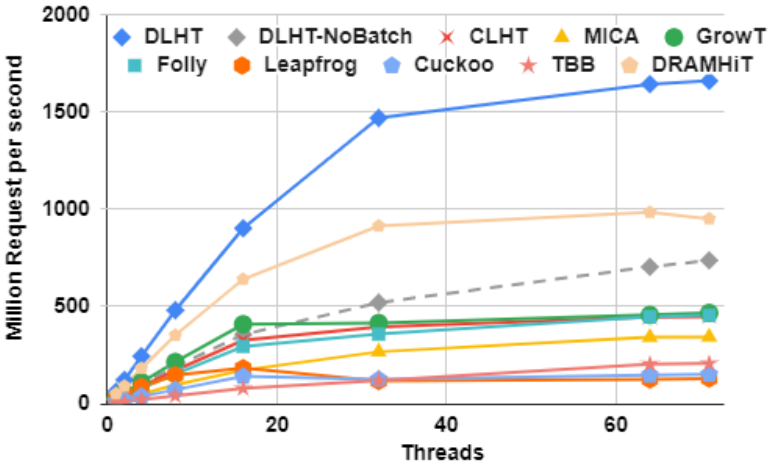}
\vspace{-22pt}
\caption{Get throughput}
\label{fig:get_ptr}
&

\includegraphics{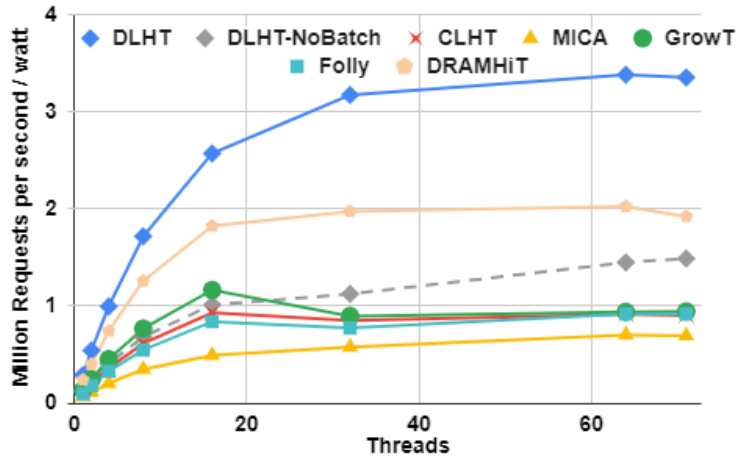}
\vspace{-22pt}
\caption{Get power-efficiency}
\label{fig:get_power}

&
\includegraphics[width=0.35\textwidth]{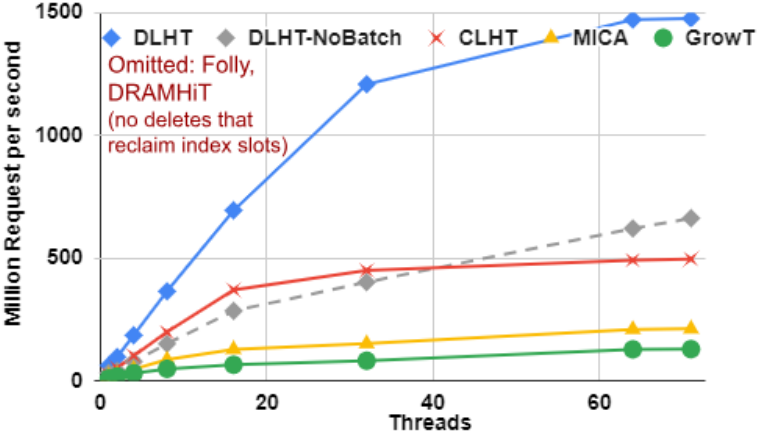}
\vspace{-22pt}
\caption{InsDel throughput 
}
\label{fig:ins_del}
\end{tabularx}
\vspace{-25pt}
\end{figure*}
By default, we run experiments using 16 threads with 8-byte keys and 8-byte values inlined in the index. We typically use a simple modulo operation for hashing. Otherwise, we use wyhash~\cite{wyhash:23}.
We preload the mimalloc allocator~\cite{mimalloc:23} and configure it to use (2MB) huge pages. 
We chose wyhash and mimalloc after extensive experimentation with the available hash functions and allocators. For a fair comparison, we use the same allocator and huge pages in all hashtables.
When threads are pinned in both sockets, we interleave the bins of the index across both sockets. 
Unless stated otherwise, we instantiate \dlht, with 67 million bins, 8.3 million chained buckets,  and we disable resizing.
Before beginning the experiment, we populate the hashtables with 100M keys.


\beginbsec{Workloads}
As shown in \tabref{tab:config}, we have two default workloads: 1)~\textit{\gets\ }with 100\% Gets and 2) \textit{\insdel\ }with 50\% Inserts and 50\% Deletes. 
To execute a Get (similarly a Put), we first select one of the prepopulated keys uniformly at random, using a random-number generator (RNG).
Inserts also use the RNG to select a key, but they choose a key that has not been prepopulated. 
This ensures that Inserts will always incur the full overhead of the insertion.
In the \insdel\ workload, an Insert is always followed by a Delete to the same key.
Details on non-default workloads are provided inline within each corresponding experimental discussion.




\begin{figure*}[t]
\setkeys{Gin}{width=1.05\linewidth}
\vspace{-3pt}
\begin{tabularx}{\linewidth}{XXX}
\hspace{-7pt}
\includegraphics[width=0.3175\textwidth]{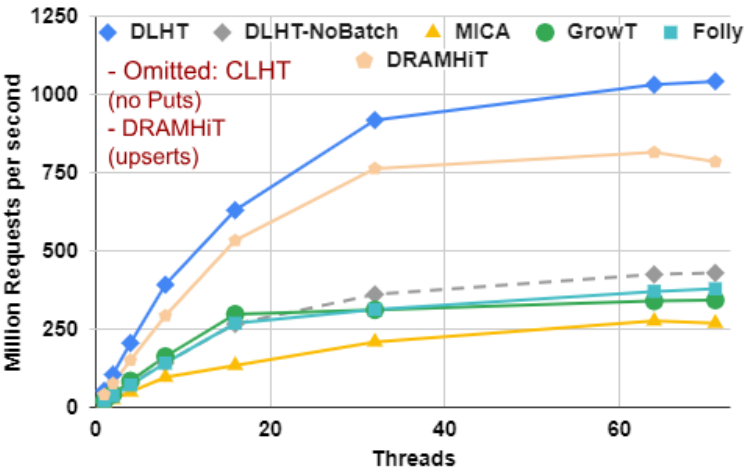}
\vspace{-10pt}
\caption{Put-heavy throughput 
}
\label{fig:put_heavy}
&

\includegraphics{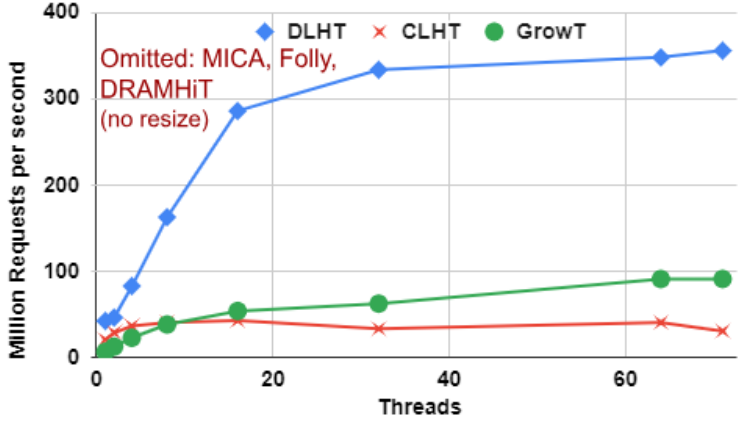}
\vspace{-22pt}
\caption{Avg. Population throughput: Inserting 800M keys over a growing index. 
}
\label{fig:population}

&
\includegraphics{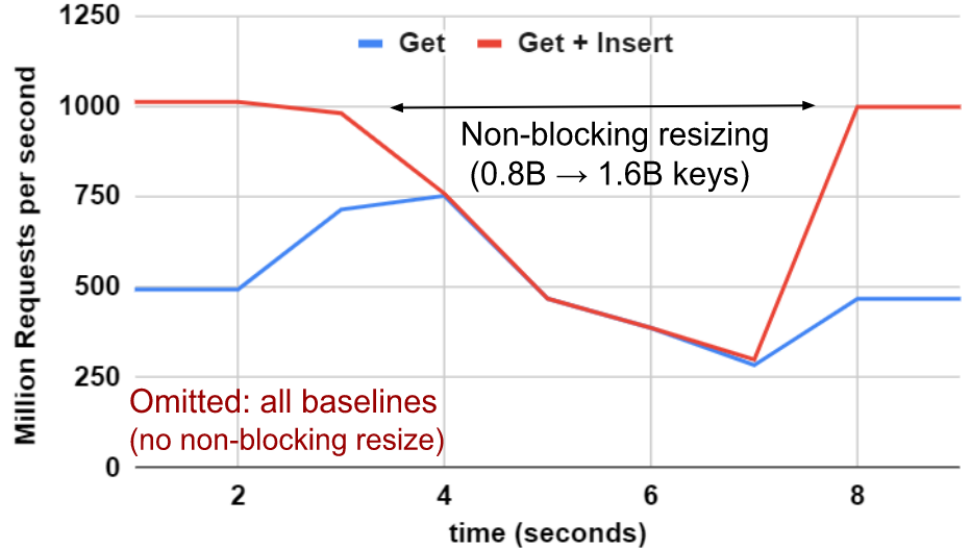}
\vspace{-22pt}
\caption{Gets, Inserts, and Non-blocking Resizing (transfer of 800M keys) in DLHT.
}
\label{fig:resizing}
\end{tabularx}
\vspace{-12pt}
\end{figure*}
\begin{table}[t]
\resizebox{1.7\linewidth}{!}{
\begin{tabularx}{2.\linewidth}{X}
\begin{tabular}{l|l}
\hline
\textbf{CLHT} & the lock-free variant of CLHT~\cite{Tudor:15, Tudor:14} \\
\textbf{MICA} & the CRCW variant of MICA2~\cite{Li:2015, Li:2016} \\
\textbf{GrowT} & the uaGrowT variant of GrowT~\cite{Maier:19} \\
\textbf{Folly} & the atomic hash map of Meta's Folly~\cite{Folly:23} \\
\textbf{Cuckoo} & the concurrent Cuckoo hash map~\cite{Fan:2013} \\
\textbf{Leapfrog} & Leapfrog map of Junction's library~\cite{Leapfrog:16} \\
\textbf{TBB} & the concurrent hash map of Intel's OneTBB~\cite{TBB:23} \\
\textbf{DRAMHiT} & a lock-free non-resizable map (concur. work)~\cite{Narayanan:23} \\
\textbf{\dlht(-NoBatch)} & as in \cref{sec:design} (without batching -- \cref{sec:des:batch})
\\ \hline
\end{tabular}
\end{tabularx}
}
\caption{Summary of evaluated Hashtables.}
\label{tab:base}
\vspace{-25pt}
\end{table}


\section{Evaluation}
\label{sec:eval}

In this section, we 
first, compare \dlht\ with state-of-the-art hashtables and test our hypotheses (\secref{sec:eval:base}). 
Then we perform sensitivity studies over \dlht\ (\secref{sec:eval:sens}) and 
investigate its performance on
benchmarks and applications (\secref{sec:eval:work}).

\subsection{Claims and comparison with state-of-the-art} 
\label{sec:eval:sota}
In the introduction, we argued for the opportunity to achieve a billion requests per second on a \emph{memory-resident} hashtable whose requests require main memory accesses. 
We claimed that state-of-the-art concurrent in-memory hashtables are unable to reach this goal 
due to excessive blocking and inefficient handling of memory accesses. Finally, we argued that state-of-the-art designs either lack or offer impractical support for core operations (i.e., Put, Delete, or Resize).
In this section, we set these hypotheses to the test.


\beginbsec{Evaluated Hashtables} In this section,
we evaluate eight state-of-the-art concurrent in-memory hashtables (baselines) and two variants of \dlht\ (shown in \cref{tab:base}). 
We 
ensure all hashtables are measured with the default configuration of \tabref{tab:config}. 
Note that almost all baselines only support up to 8-byte values.
Therefore,
we use 8-byte values in this subsection. We extensively explore different sizes and configurations of \dlht\ in the subsection with sensitivity studies (\cref{sec:eval:sens}).



For simplicity, we mainly focus on comparing with GrowT, \linebreak DRAMHiT, Folly, CLHT, and MICA, as these are the fastest baselines. We analyzed the features of these hashtables in \cref{tab:core_features}. In short, GrowT, DRAMHiT, and Folly are
open-addressing designs.
Folly does not support Resizes or Deletes with slot reclamation. \growt\ offers parallel but blocking Resizes and impractical Deletes that must perform a blocking transfer of all objects to a new index to reclaim index slots.
The lock-free variant of CLHT, assumes unique values, does not support Puts, and cannot chain buckets. The latter limits collisions to one cache-line (i.e., three objects), after which it performs a Resize that problematically is blocking and single-threaded.
MICA uses software prefetching for its memory accesses but is lock-based, lacks support for Resizes, and cannot inline keys and values inside its index. Thus, MICA needs at least two memory accesses to access or update a value and incurs (de)allocation overheads on all (Deletes) Inserts. 
Finally, DRAMHiT is the only design that combines frugal memory accesses and prefetching like DLHT. However, DRAMHiT lacks Resizing, its Deletes cannot reclaim index slots, its Puts (Inserts) may silently insert (update) key-values, and its batching may reorder a client's requests.

\label{sec:eval:base} 
\subsubsection{Get throughput and power-efficiency} \label{sec:eval:get}

\cref{fig:get_ptr} illustrates the throughput of the Get workload in million requests per second (\mreqs) as we vary the number of threads. Our \dlht\ implementation outperforms all baselines, offers beyond 1B Gets/s, and scales almost linearly up to 32 cores across both available sockets. With hyper-threading enabled (i.e., 64 and 71 threads), \dlht\ saturates the memory bandwidth and peaks at 1.66B Gets/s. Its high performance mainly stems from serving each Get with a single, prefetched memory access and no write-backs.

In contrast, (\dlht-NoBatch,) GrowT, Folly, and CLHT  can serve Gets with a single memory access but do not leverage prefetching. Thus, they are over (2.2$\times$) 3.5$\times$ slower than \dlht. Similarly, MICA uses prefetching, but its non-inlined approach requires at least two memory accesses for every Get, resulting in up to 4.8$\times$ lower throughput than \dlht. 
DRAMHiT, which combines an inlined index with software prefetching, is only 1.7$\times$ slower than \dlht.
However, unlike \dlht, DRAMHiT may reorder the execution of requests inside a client's batch.
We partially attribute DRAMHiT's lower performance to the design that does not let application threads to directly access the hashtable and the overheads of its asynchronous engine (e.g., producer-consumer queues). 

%
Finally, Cuckoo, TBB, and Leapfrog mandate more than one memory access and do not use prefetching; hence their throughput is below 250 \mreqs. For simplicity, we omit those baselines from the rest of our graphs and focus our comparisons on 
GrowT, DRAMHiT, Folly, CLHT, and MICA.

\begin{figure*}[t]
\setkeys{Gin}{width=1.\linewidth}
\begin{tabularx}{\linewidth}{XXXX}
\includegraphics{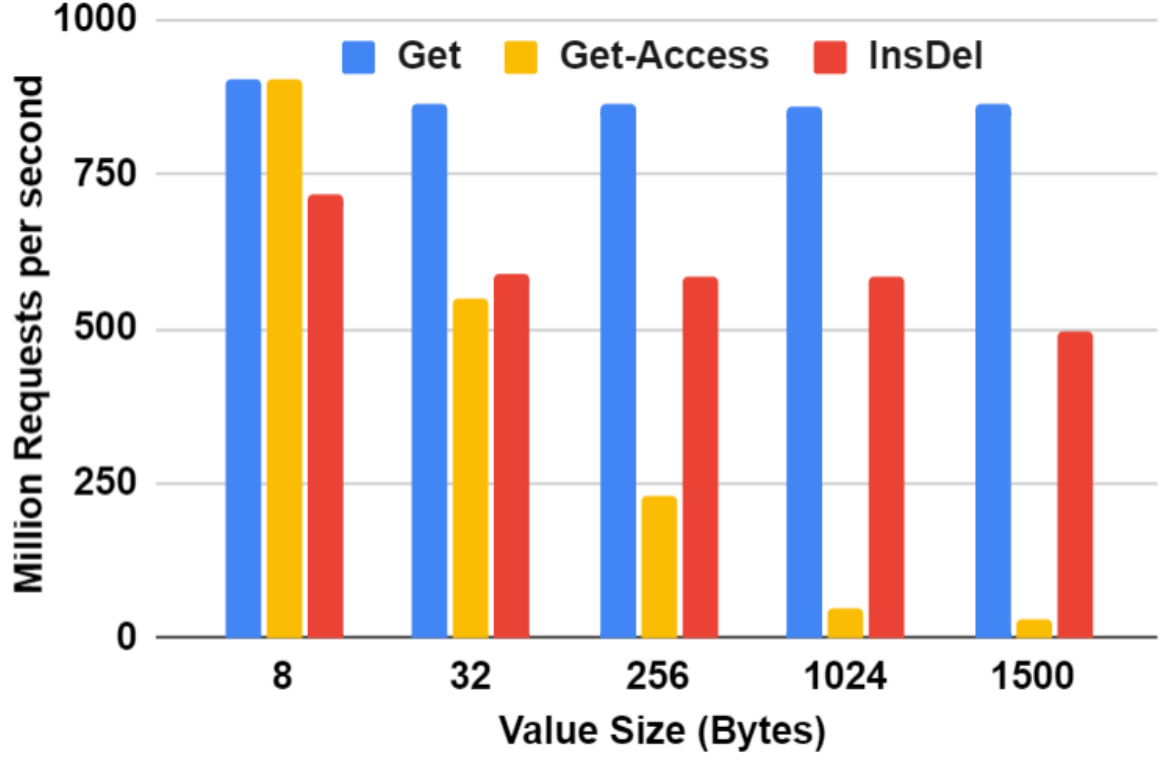}
\vspace{-22pt}
\caption{Varying value size}
\label{fig:value_size}
&

\includegraphics{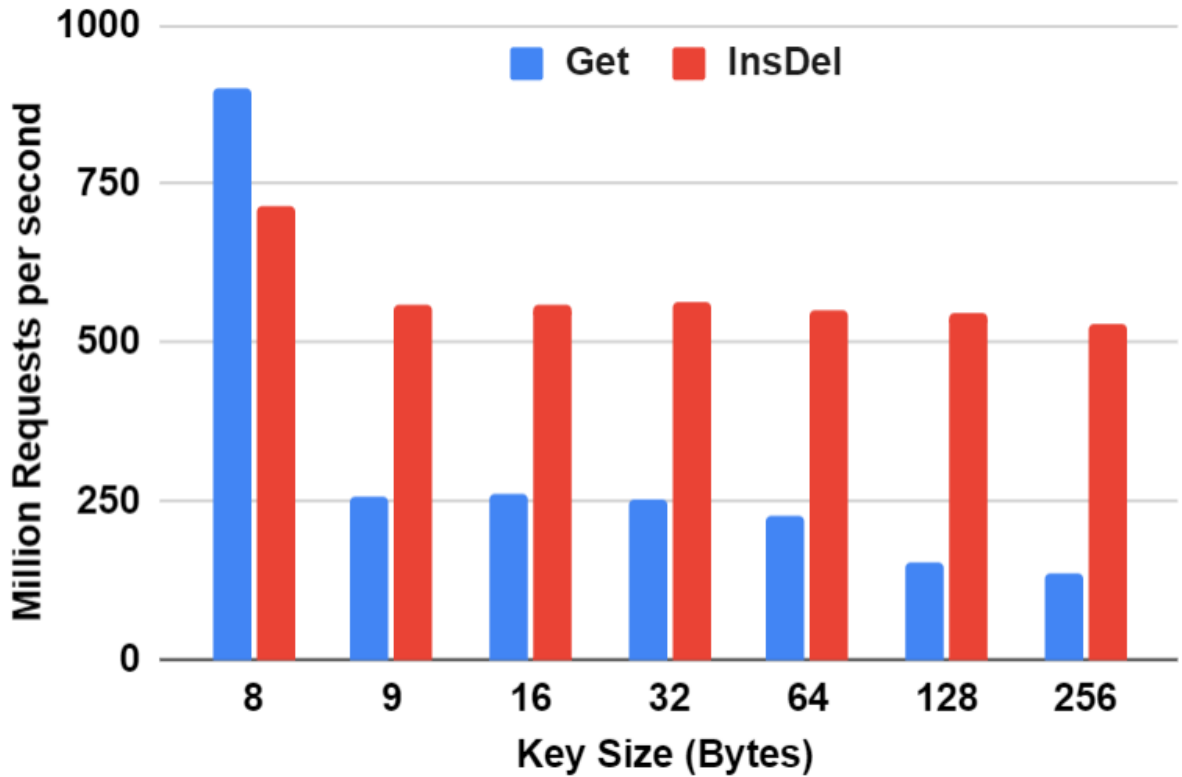}
\vspace{-22pt}
\caption{Varying key size}
\label{fig:key_size}
&

\includegraphics{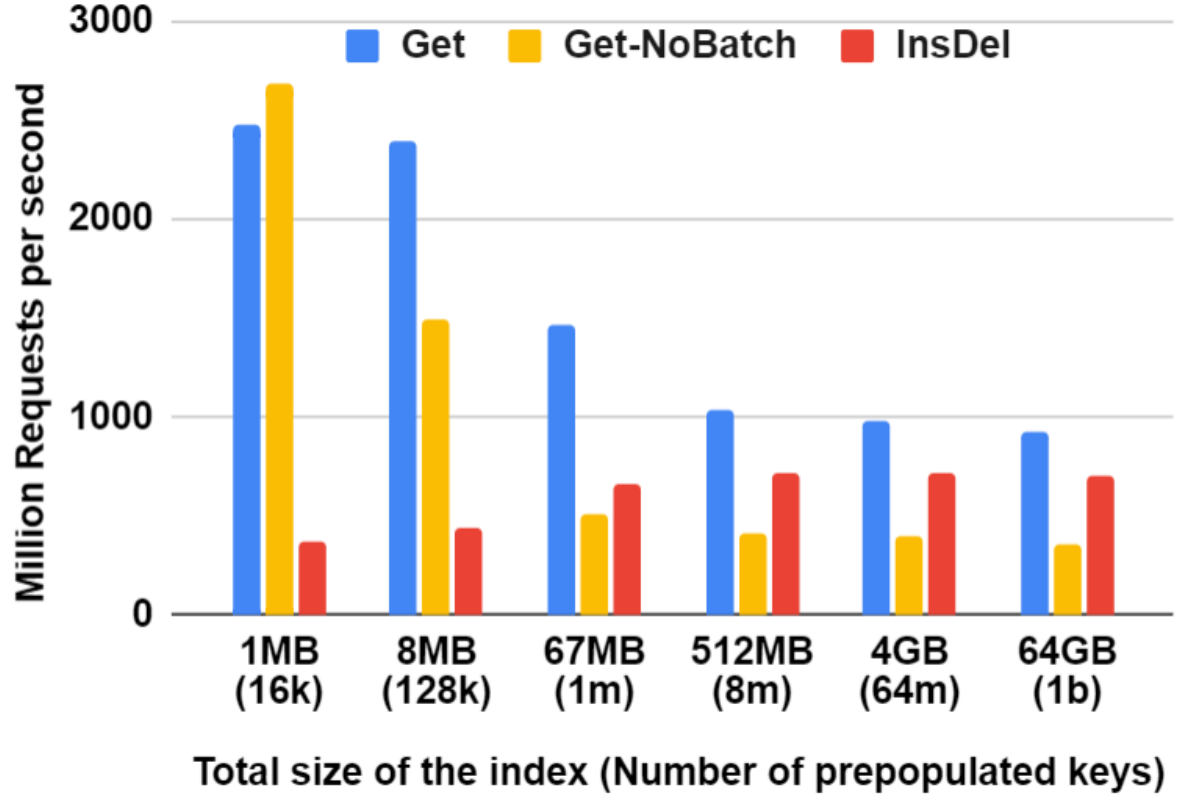}
\vspace{-22pt}
\caption{Varying index size}
\label{fig:key_num}


\end{tabularx}
\end{figure*}
\cref{fig:get_power}, shows the power efficiency on the Get workload in \mreqs\ per watt. The power efficiency of \dlht\ sees a steady increase up to 32 threads and peaks at 3.35\mreqs\ per watt, which is up to  1.7$\times$ higher than (the non-resizable) DRAMHiT and 3.6$\times$ higher than (the resizable) GrowT, the two most efficient baselines on Gets.
In short, \dlht\ fulfills its purpose. It offers substantially higher throughput 
than the competition, surpassing 1.6B requests per second with its memory-access-aware design without sacrificing functionality. Next, we focus on functionality, starting with Deletes, which is a key point of criticism for open-addressing hashtables.




\subsubsection{Delete throughput}\label{sec:eval:insdel}
Deleting elements in a concurrent open-addressing scheme is cumbersome. 
In fact, neither DRAMHiT nor Folly support Deletes that free index slots.
In this section, we investigate Deletes in open-addressing and contrast them with \dlht.

\growt\ supports Deletes through tombstones. As the client is issuing Deletes, over time, the index fills with tombstones. When the index gets filled beyond a threshold, \growt\ creates a new index, where it must move all the alive keys. We perform an experiment with the common pattern of an Insert followed by a Delete to the same key (i.e., the default InsDel workload). We start with an empty hashtable that can fit 100 million keys. \figref{fig:ins_del} shows the completed Inserts and Deletes in \mreqs\ for increasing thread counts.

\dlht\ achieves up to 12.8$\times$ higher throughput than \growt. This is because \growt\ must move to a new index roughly every 100 million Deletes.
Notably,
this is still a favorable experiment for \growt, because we allocate a very large table compared to the number of alive keys to amortize the cost of moving to a new table every 100M Deletes, and when it needs to move to a new table, only a handful of keys are actually moved. The latter is because there is always only at most one alive key per thread at any given moment.

The closed-addressing but inlined incapable scheme of MICA also suffers in this workload. This is because MICA requires at least two memory accesses and two write-backs to memory for each Insert and Delete to a key. As expected, CLHT performs comparably to 
\dlht-NoBatch, as both perform the Insert and Delete to the object via accessing and updating the same single cache-line. On top of that, \dlht\ with prefetching overlaps the blocking of memory accesses, hence achieving up to almost 3$\times$ the throughput of CLHT.

\subsubsection{Put throughput} \label{sec:eval:putheavy}
In \cref{fig:put_heavy}, we evaluate a Put-heavy workload with 50\% Gets and 50\% Puts over the default configuration of \cref{tab:config} and while varying threads. 
\dlht\ reaches up to 1042 \mreqs\ and outperforms both open-addressing baselines that do not utilize software pre\-fetching up to 2.7$\times$. 
DLHT provides a smaller benefit over DRAMHiT, which also exploits prefetching. However, Puts in DRAMHiT may silently insert an item in the index and may be performed out-of-order.
%
As in the InsDel experiment, MICA requires multiple memory accesses for every Get and Put hence cannot maximize performance despite utilizing prefetching.
Finally, recall that CLHT does not support Puts 
and is thus omitted from the graph.

\subsubsection{Insert and Resizing throughput} \label{sec:eval:resize}
To evaluate \dlht's non-blocking resizing, we perform two experiments. First, we populate 800 million keys in an initially small index that grows on demand. \cref{fig:population} shows the throughput as the thread count increases and includes the baselines that support resizing.
As expected, we observe that inserting in a growing index is detrimental to CLHT, which cannot increase its population throughput beyond 8 threads, as the cost of its single-threaded (blocking) resize dominates the population.
GrowT's population with parallel resizing scales better with threads, but its blocking nature and slower Inserts render GrowT's throughput up to 3.9$\times$ lower than the parallel non-blocking population of \dlht.





To better demonstrate the non-blocking nature of \dlht's resize, we run a second experiment where 32 threads populate the \dlht\ hashtable to 1.6B keys and 32 additional threads randomly perform Gets on these keys. 
\cref{fig:resizing} shows the throughput of Gets and Inserts over time and a full index of 800M keys that starts resizing (to a 1.6B keys index) at approximately 3 seconds.
At the early stages of the resize, threads complete Inserts as long as they still find space in the old index. Once they cannot fit their object into the old index, they help the resizer to accelerate the transfer. 
Helping allows this large transfer to complete in about 4 seconds.
For comparison, it takes GrowT (CLHT) more than 5.6 (14) seconds to complete the same transfer in a parallel (sequential) but blocking way, during which all operations are stalled.

In contrast, the non-blocking resize of \dlht\ allows other operations (Gets in this case) to complete safely without waiting. Gets during a resize that find their bin of the old index in a state other than DoneTransfer, proceed as normal with a single (prefetched) memory access. As bins are transferred, more Gets 
must pay the overhead of accessing both the old and the new index until the resize completes. Hence, the Get throughput degrades over time during transferring but comes back once the transfer completes.
%
%

\subsubsection{Index occupancy}
\label{sec:eval:occupancy}
To study the occupancy, we use wyhash and populate a growing index.
We limit the number of link buckets to one-fifth of the bins to keep serving most operations with a single memory access. Recall that the lock-free CLHT does not support chaining. Meanwhile, DLHT chains up to 3 extra buckets per bin. Not surprisingly, even with a state-of-the-art hash function, we observe that CLHT's inability to chain buckets results in low occupancy (1-5\%). In contrast, DLHT gracefully handles collisions via its bounded chaining and sees 63-72\% occupancy. 
Note that to maintain good performance, open-addressing hashtables typically resize when they reach 30-50\% occupancy (30\% in GrowT's codebase)~\cite{Maier:19, Growt:23}.


\subsection{Sensitivity studies on \dlht}\label{sec:eval:sens}
In this section, we isolate our focus on \dlht\ to characterize its performance while varying its features.

\begin{figure*}[t]
\setkeys{Gin}{width=1.\linewidth}
\begin{tabularx}{\linewidth}{XXX}

\includegraphics{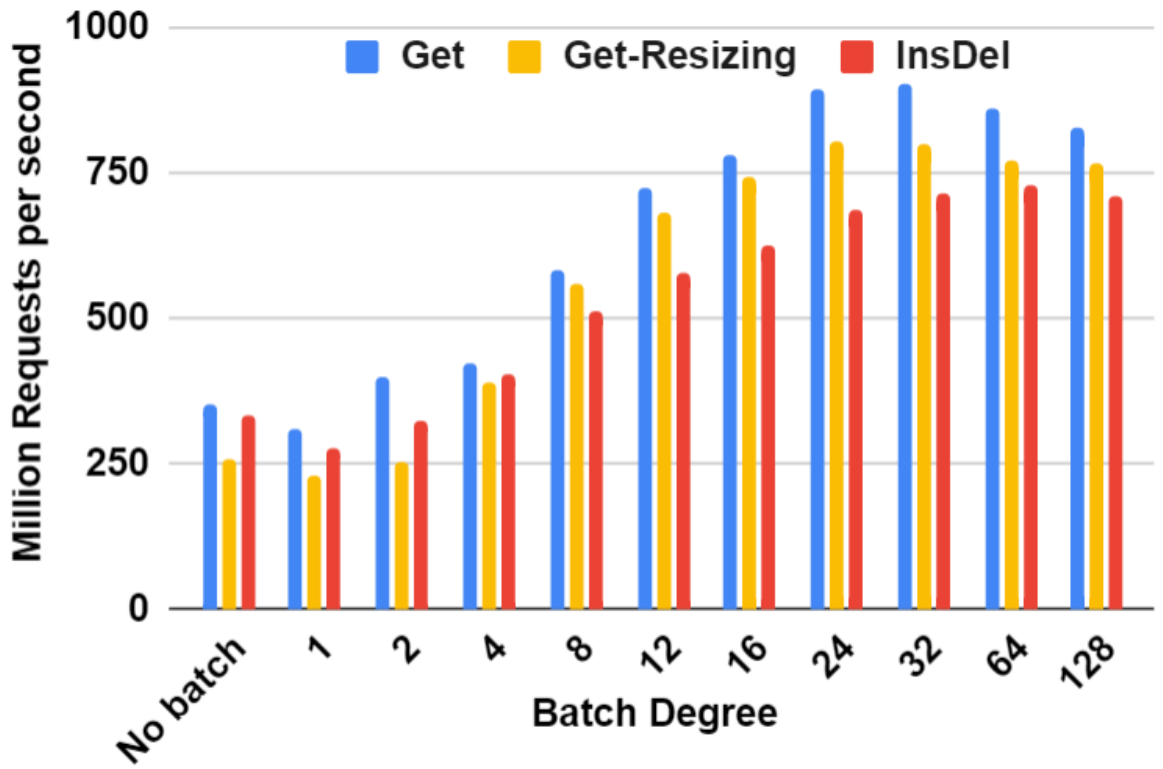}
\vspace{-22pt}
\caption{Varying batch size}
\label{fig:batch}
&

\includegraphics{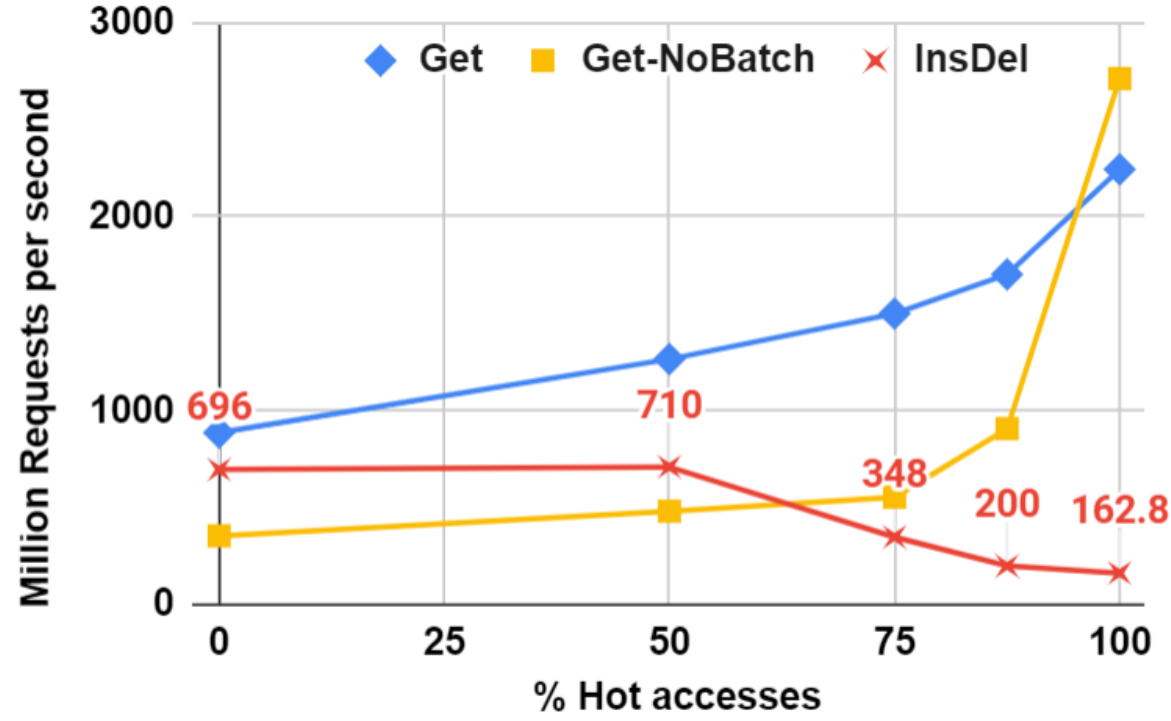}
\vspace{-22pt}
\caption{Skew with 1000 hot keys}
\label{fig:skew}
&

\includegraphics{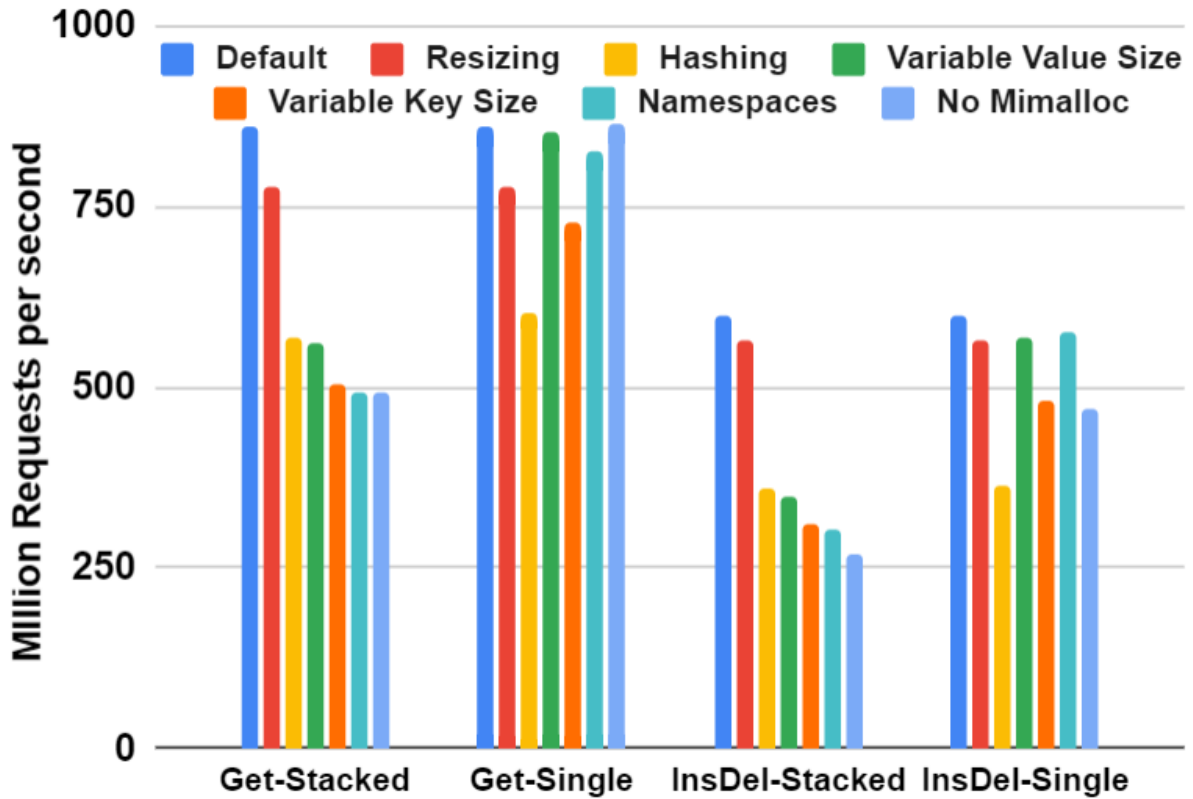}
\vspace{-22pt}
\caption{Enabling \dlht\ features}
\label{fig:overheads}
\end{tabularx}
\end{figure*}

\subsubsection{Varying value size and key size}\label{sec:eval:var_val_key}
In \figref{fig:value_size}, we vary the value size from 8 to 1500 bytes. When values are 8 bytes, they are inlined in the index, and we do not need to allocate any extra space for them.
Apart from our two default workloads (\gets\ and \insdel), we also run a workload called Get-Access. 
This is different from the \gets\ workload as after every Get, we read the whole value. Recall that when the value is larger than 8 bytes, Gets only return a pointer to the value without accessing it.

For this reason, the \gets\ workload is only slightly affected as the value size increases.
The throughput of \insdel\ gradually decreases as the size of allocations increases.
The throughput of Get-Access drops quickly as each get is associated with extra memory accesses to read the entire value from memory.

In \figref{fig:key_size}, we run the two default workloads while varying the key size from 8 to 256 bytes. 
Note the steep performance drop when the key is larger than 8 bytes.
In this case, the key
does not fit in the bucket slot and is stored with the value. This means that 1) all Gets must now dereference the pointer to the value to read the entire key (an 8-byte signature is left on the bucket for filtering), 2) we need to allocate more bytes per Insert to fit the key in the value, and 3) each Insert must write the key to the newly allocated value.

\subsubsection{Varying index size}
We next vary the index size of \dlht\ from 1MB (with 8K prepopulated keys) to 64GB (with 1B prepopulated keys). 
To study when prefetching is beneficial,
apart from the two default workloads (\gets\ and \insdel), we run a workload called Get-NoBatch, which is the same as \gets\ but without batching. 

We make two observations. First, because the smallest index (1MB) can fit inside the private L2 of each core, prefetching is not helpful, and thus batching has no benefit, only overheads. As the index grows, prefetching, and thus batching, becomes increasingly beneficial. 
Second, we see that unlike \gets, \insdel\ does not benefit from a small cache-resident index. 
This is because a smaller index increases the chances of conflicts in the same bin. This affects the performance as Inserts and Deletes must CAS the header of the bin.


\begin{figure*}[t]
\setkeys{Gin}{width=1.0\linewidth}
\begin{tabularx}{\linewidth}{XXX}

\includegraphics{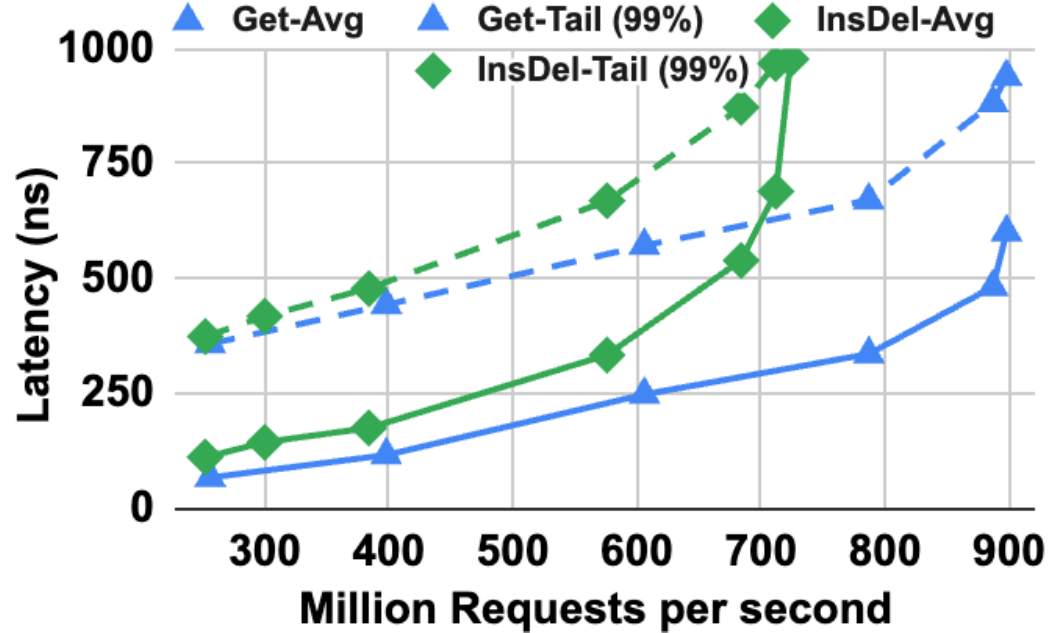}
\vspace{-22pt}
\caption{Latency of Gets and InsDel}
\label{fig:latency_both}
&



\includegraphics{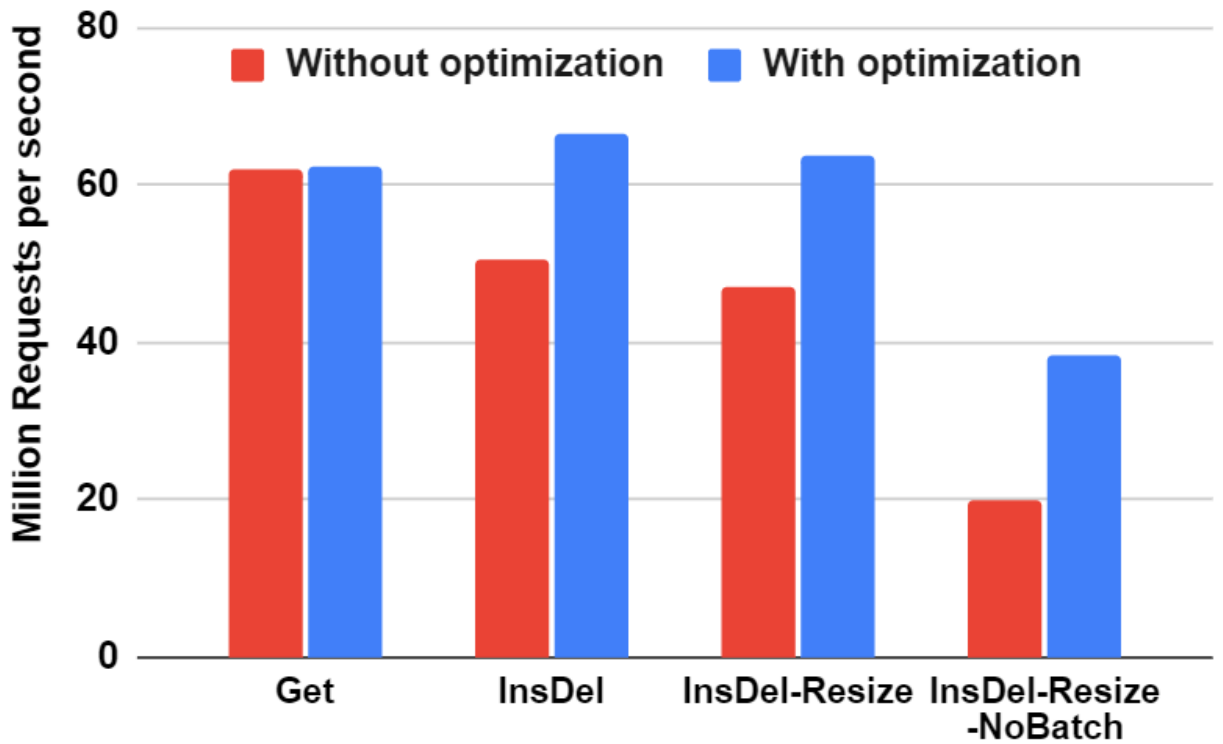}
\vspace{-22pt}
\caption{Single thread application}
\label{fig:single_thread}
&

\includegraphics{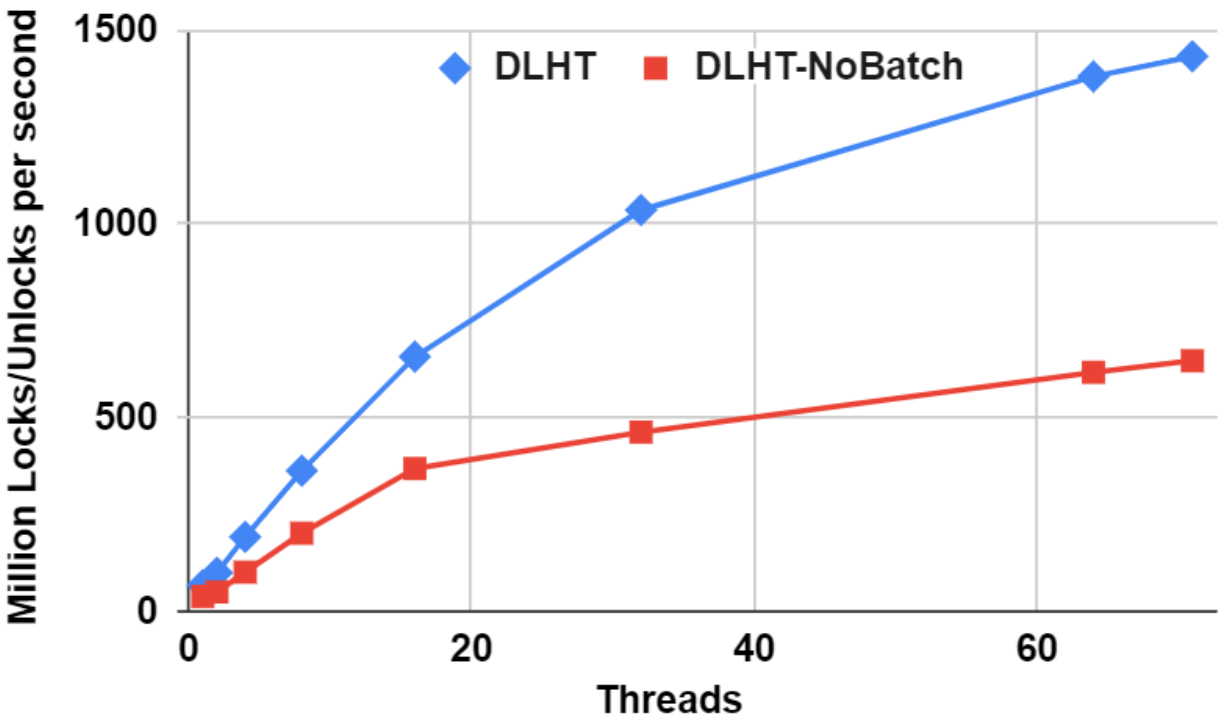}
\vspace{-22pt}
\caption{Lock Manager}
\label{fig:lock_unlock}

\end{tabularx}
\vspace{-20pt}
\end{figure*}
\subsubsection{Varying batch size}
Batching is a key feature of \dlht\ where software prefetching is used to overlap 
memory accesses for a request with useful work on other requests.
\cref{fig:batch} shows the throughput of \dlht\ under the default configuration, while varying the batch degree, \ie the number of requests grouped in a batch.
Apart from our two default workloads (\gets\ and \insdel), we also show a workload called \gets-Resizing, which is the same as \gets\ but with resizing enabled. This does not mean that the workload resizes the index, but rather that we have compiled \dlht\ with the capability to resize, if needed. The bars denoted as \qt{No batch}, do not use the batching API.

We make three observations. First, batching can be very effective in boosting performance.
The gains saturate around 24 accesses; we attribute this limitation to hardware (number of MSHRs and TLB size). Second, batching improves throughput if there are at least four requests (two for \gets) that can be overlapped. Otherwise, the overhead of creating the request array trumps the benefit of overlapping memory accesses. Third, enabling resizing hurts performance more when no batching is used. This is because resize requires each request to perform two additional atomic stores (\cref{sec:des:alg:resize}). Batching amortizes this overhead amongst the batched requests.


\subsubsection{Skew}
To study the behavior of \dlht\ under skew, we perform an experiment with the default configuration, but 1000 keys are hot and receive an increasing percentage of accesses. \cref{fig:skew} shows the throughput of \dlht\ for the \gets, \insdel, and Get-NoBatch while we increase the percentage of skewed accesses. As expected, in Get and Get-NoBatch the throughput increases with skew, surpassing 2.2B Gets/s as both benefit from the locality of skewed accesses. When all accesses target hot keys, prefetching stops being useful and Get-NoBatch outperforms the Get workload that includes the overheads of batching. 
Under high skew, InsDel manifests a high number of conflicts which impacts its performance.

\subsubsection{Enabling features}
\dlht\ has several features that can be enabled at runtime or compile-time. Enabling these features typically incurs a performance penalty. \cref{fig:overheads} shows the throughput in \mreqs\ of the two default workloads (\gets\ and \insdel) while enabling features either in a \emph{stacked} fashion, where features are enabled on top of each other, or one-at-a-time (i.e., \emph{single}). 
We explain the meaning of each bar below. 

\beginbsecnospace{Default}This bar shows the throughput of the default configuration (\tabref{tab:config}), with the exception that we use 32-byte values. This serves as the baseline for the rest of the bars.

\beginbsecnospace{Resizing} \dlht\ is compiled with resizing enabled. Note that no resizing actually takes place in this experiment, but \dlht\ incurs a performance penalty, as each thread must notify other threads when entering and leaving the \dlht\ index, as well as check if other threads have initiated resizing. 

\beginbsecnospace{Hashing}
\dlht\ uses the wyhash hash function~\cite{wyhash:23} instead of a simple modulo operation to hash a key to a bin. Naturally, this also further reduces throughput. 

\beginbsecnospace{Variable value size}
\dlht\ is compiled to allow every value to have a different size. The performance penalty is very small because the only overhead is that the size of each value must be stored in the header of the value.

\begin{figure*}[t]
\setkeys{Gin}{width=1.0\linewidth}
\begin{tabularx}{\linewidth}{XXX}


\includegraphics{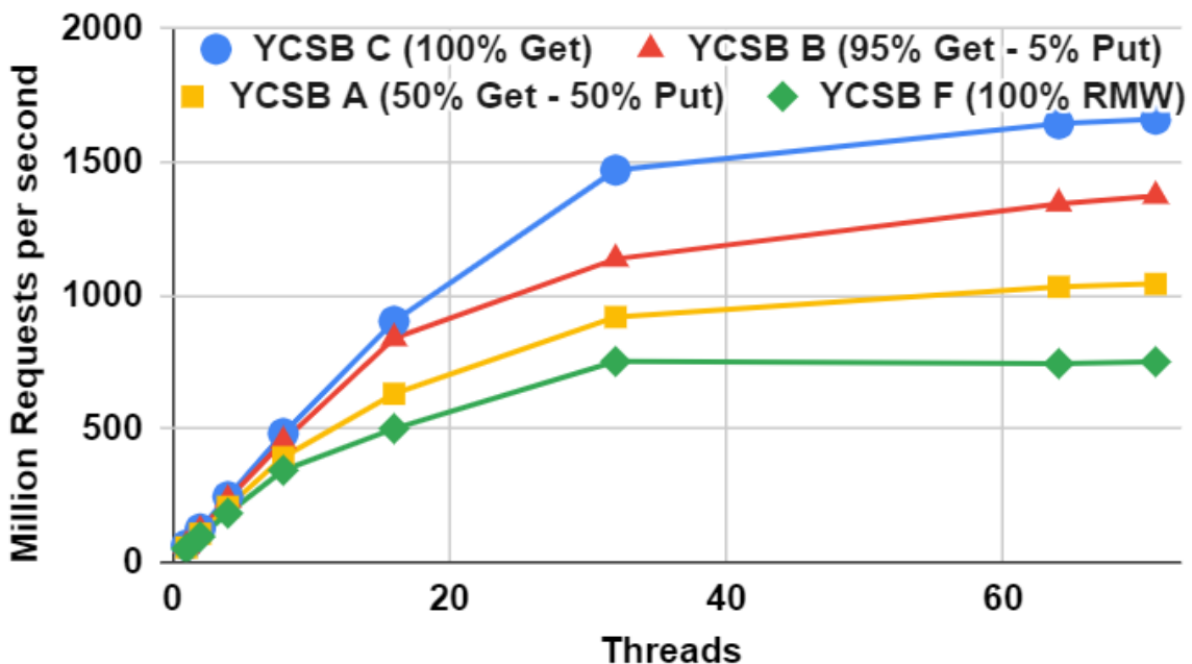}
\vspace{-22pt}
\caption{Single-key YCSB mixes}
\label{fig:ycsb}
& 

\includegraphics[width=0.30\textwidth]{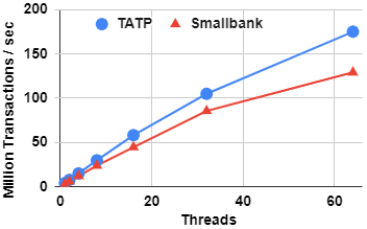}
\vspace{-12pt}
\caption{Transactional benchmarks}
\label{fig:txs}
&

\includegraphics[width=0.33\textwidth]{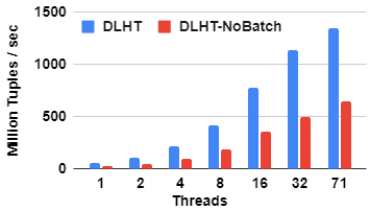}
\vspace{-22pt}
\caption{Hash Joins}
\vspace{-15pt}
\label{fig:join}

\end{tabularx}
\vspace{-10pt}
\end{figure*}

\beginbsecnospace{Variable key size}
\dlht\ allows every key to have a different size. In this experiment, all used keys are smaller than 8 bytes. Hence the only overhead is storing and checking the key size in the most significant bits of the pointer (discussed in \secref{sec:des:add:var}).
The overheads for larger keys are studied in \secref{sec:eval:var_val_key}

\beginbsecnospace{Namespaces}
\dlht\ is compiled with namespaces, incurring a small penalty, as the namespace is stored and checked in each search in the most significant bits of the value pointer. 

\beginbsecnospace{No mimalloc}
Finally, we use the standard library malloc instead of mimalloc. Naturally, this impacts only the \insdel\ workload, which calls either malloc or free on every request.

Notably, this is far from an exhaustive evaluation of all possible combinations of runtime and compile-time options, which would have been intractable to evaluate. 
Rather it is a selection of popular features among our clients.

\subsubsection{Latency}
While we focus on delivering high throughput, we also study the latency of \dlht\ in this section. \cref{fig:latency_both} shows the average and tail (99th\%) latency in nanoseconds for our Get and InsDel workloads, as a function of the load. Unsurprisingly, the nature of memory-resident workloads translates into average latencies of 100s nanoseconds which increase with the load. We also observe that the tail is below a microsecond, even under high load. Naturally, Get has lower latencies than InsDel which involves slower CAS operations.

\subsection{Applications and Benchmarks over \dlht}\label{sec:eval:work}
In this section, we evaluate applications and benchmarks that represent some of our target use-cases. 

\subsubsection{Optimized single-threaded application} 

We optimized \dlht\ to deliver high performance on single-threaded applications.
\cref{fig:single_thread} shows the single-threaded throughput with and without our optimizations under four workloads. 
Apart from our two default workloads, we also run 
\insdel-Resize, which enables resizing, and \insdel-Resize-Nobatch, which enables resizing but disables batching.

Starting from \insdel, we see a 31$\%$ improvement as a concurrent Insert requires two CASes, and a Delete requires one CAS. With the optimization, we convert these into regular stores. \insdel-resizing sees a 35$\%$ improvement, as resizing requires notifying other threads when entering and leaving \dlht, via atomic stores. The optimization removes the need to notify other threads, as there are no other threads. The impact of this is amplified when batching is disabled (91$\%$ improvement), as the enter/exit notification must now happen once per request instead of once per batch.
Finally, we note that the optimization has no impact on the \gets\ workload because 
on x86 8-byte atomic loads have a trivial overhead.

\subsubsection{Remote memory: CXL emulation}
Memory latency seldom improves~\cite{Wulf:95}. In fact, recent trends increase capacity over cost by introducing slower memories that are denser (e.g., NVM) or reside far (e.g., as in RDMA or CXL). 
Hence, we believe that memory-aware designs, which minimize accesses and exploit software prefetching, will remain critical in next-generation deployments. 

To test this hypothesis, we 
run the Get workload while emulating 
a CXL deployment of DLHT. For this, we run 36 threads on the first socket and pin DLHT's memory on the remote socket -- as in recent works~\cite{Li:23, Maruf:23}.
We observe that DLHT (with prefetching) offers 2.9$\times$ the throughput of DLHT-NoBatch. As expected, remote memory lowers the throughput (to about half) compared to local.

\subsubsection{Lock manager over HashSet}
\label{sec:eval:lockm}
This workload is inspired by a client that used \dlht's HashSet to implement a database lock manager. Inserting and deleting a key over the HashSet mimics locking and unlocking a key (or a set of keys). 
Recall that \dlht's batching interface preserves the order of requests.
As such, \dlht's order-preserving batching is an especially good fit for a lock manager, which typically wants to implement protocols similar to two-phase-locking that grab locks in some order and later release them. 
This is also an application example where DRAMHiT's out-of-order batching, which may reorder requests to overlap their accesses, can introduce deadlocks.

\cref{fig:lock_unlock} shows the throughput of locks and unlocks 
while varying the threads for \dlht(-NoBatch). When batching is used, the throughput scales well with the number of threads and is able to peak at almost 1.5B locks/unlocks per second. As expected without batching, when the memory accesses cannot be masked, the throughput is up to 2.2$\times$ lower.

\subsubsection{Single-key benchmarks (YCSB Mixes)}
We study the performance of \dlht\ under YCSB~\cite{Cooper:2010}, the popular single-key benchmark. \cref{fig:ycsb} shows the throughput of four YCSB mixes (shown in legend) as the threads increase. The configuration follows the defaults of \cref{tab:config}. 

All four mixes scale well up to 32 threads where cores in both sockets are used and do saturate the memory bandwidth when hyper-threading is enabled. The update-only (YCSB F) mix peaks at about half the throughput of read-only (YCSB C), saturating the memory bandwidth at 32 threads. This is expected, since it incurs double the memory traffic as every accessed cache-line is updated and written back to memory. 
\subsubsection{Multi-key benchmarks (OLTP Transactions)}
To study the performance of \dlht\ on multi-key transactions, we implement and evaluate two popular OLTP benchmarks, TATP~\cite{TATP:2009}, which is read-intensive, and Smallbank~\cite{Smallbank:2009} that is write-intensive
(summarized in \cref{tab:bench}). In this experiment, the same threads run both the logic that generates transactions and the \dlht\ logic. We ensure the benchmarks are memory-resident by populating their tables with 1M subscribers (TATP) and 10M accounts (Smallbank).

\begin{table}[t!]

\resizebox{\columnwidth}{!}{
\begin{tabular}{l|lrrrr}
 &
  \textbf{Characteristic} &
  \multicolumn{1}{l}{\textbf{Tables}} &
  \multicolumn{1}{l}{\textbf{Columns}} &
  \multicolumn{1}{l}{\textbf{Txs}} &
  \multicolumn{1}{l}{\textbf{Read Txs}} \\ \cline{2-6} 
\textbf{TATP}      & read-intensive  & 4 & 51 & 7 & 80\% \\
\textbf{Smallbank} & write-intensive & 3 & 6  & 6 & 15\% \\
\end{tabular}
}

\vspace{5pt}
\caption{Summary of evaluated transactional benchmarks.}
\vspace{-15pt}
\label{tab:bench}
\end{table}

\cref{fig:txs}, shows the throughput of the two benchmarks while varying threads.
We see that both benchmarks scale well with the number of threads and surpass 100M transactions per second. With 64 threads TATP (Smallbank) reaches 175M (129M) transactions per second. Note that both benchmarks only access and/or update just a few objects in a transaction. Thus, TATP, with fewer updates, outperforms Smallbank as it incurs fewer write-backs to memory. 

\subsubsection{Hash Join (OLAP Application)}
We evaluate a non-partitioned join over DLHT. We use \textit{workload A} from~\cite{Lutz:20} in which both relations are memory-resident. Workload A consists of 16B tuples 
and a build relation ($R$) and probe relation ($S$) with size $2^{27}$ and $2^{31}$ similarly to~\cite{Schuh:16, Lutz:20}. 
\cref{fig:join}, shows
the throughput (\scalebox{0.7}{$\dfrac{|R| + |S|}{runtime}$}) while varying the number of threads. 

In joins, where batching is naturally applicable, DLHT (with software prefetching) achieves up to 1.4B tuples/sec. This translates to a $2.2\times$ benefit over DLHT-NoBatch.
Interestingly, the memory-aware and non-blocking design of DLHT delivers a very high throughput CPU-based join -- despite large tables that cannot fit in cache and under no partitioning or specializations for joins (e.g., eschewing synchronization for the read-only probing phase). We note that partitioning and other such optimizations are synergistic to the features of \dlht. We leave a more elaborate study for future work, as joins are not the main focus of this paper.

\begin{table}[t]
\resizebox{1.1\linewidth}{!}{
\begin{tabularx}{1.5\linewidth}{X}
\hspace{5pt}
\begin{tabular}{|l|l|}
\hline \colorhl
\begin{tabular}[c]{@{}l@{}}  \textbf{Baseline} \end{tabular} &
\colorhl   \textbf{Comparison with \dlht}   \\
\hline
\hline

\colorhl \begin{tabular}[c]{@{}l@{}}  \textbf{CLHT} \end{tabular} &
\begin{tabular}[c]{@{}l@{}} - 3.5x lower Get throughput | 8x slower Population \\ 
- more than 10x worse occupancy (cannot chain buckets) \\
- Blocking resize, no Puts, assumes unique values, $\leq$ 8B keys/values \end{tabular} \\ \hline \hline

\colorhl \begin{tabular}[c]{@{}l@{}}  \textbf{MICA} \end{tabular} &
\begin{tabular}[c]{@{}l@{}} - 4.8x lower Get throughput \\ 
- Non-resizable, lock-based, non-inlining \end{tabular}  \\  \hline \hline

\colorhl \begin{tabular}[c]{@{}l@{}}  \textbf{GrowT} \end{tabular} &
\begin{tabular}[c]{@{}l@{}} - 3.5x lower Get throughput | 3.9x slower Population\\ - 12.8x lower \insdel\ throughput (tombstone-based deletes) \end{tabular}  \\  \hline \hline

\colorhl \begin{tabular}[c]{@{}l@{}}  \textbf{Folly} \end{tabular} &
\begin{tabular}[c]{@{}l@{}} - 3.5x lower Get throughput \\ 
- Non-resizable, deletes cannot reclaim slots, $\leq$ 8B keys/values only
\end{tabular}  \\  \hline \hline

\colorhl \begin{tabular}[c]{@{}l@{}}  \textbf{DRAMHiT} 
\end{tabular} &
\begin{tabular}[c]{@{}l@{}} - 1.7x lower Get throughput 
\\ - Non-resizable, deletes cannot reclaim slots, $\leq$ 8B keys/values only
\\ - Only Upserts (no pure Insert/Put), batching may reorder requests
\end{tabular}  \\  \hline


\end{tabular}
\end{tabularx}
}
\vspace{5pt}
\caption{\scalebox{0.96}{Comparison summary of \dlht\ and (fastest) baselines.}}
\label{tab:eval:comp}
\vspace{-20pt}
\end{table}

\section{Related Work}
\label{sec:related-work}
This section extends related work beyond the designs summarized in~\cref{tab:eval:comp}, which we also analyzed in \cref{sec:background} and \cref{sec:eval:sota}.

Numerous works focus on high-throughput hashtables.
Some exploit hardware acceleration via FPGAs~\cite{Chalamalasetti:13, Lavasani:14, Istvan:15, Li:2017}, GPUs~\cite{Hetherington:15, Zhang:15, Breslow:16, Lutz:22, Lutz:20, Schuh:16}, or SIMD~\cite{Folly:23, Abseil:23}. 
%
Others propose alternative probing~\cite{Shalev:06, Fan:2013, Dragojevic:2014}, efficient concurrency control~\cite{TBB:23, Lim:2014}, or
focus on persistency~\cite{Hu:2021, Nam:19, Lee:19, Zhao:23, Lu:20V, Chen:20, Chandramouli:18}.
Finally, some designs offer non-blocking resizes~\cite{Gao:04, Liu:14, Junchang:20, Greenwald:02, Fatourou:18}.
Unlike \dlht,
none of 
these designs 
exploit software prefetching to hide latencies and most need two or more memory accesses to serve a request from memory.
\section{Conclusion}
\label{sec:conclusion}
We showed that existing in-memory hashtables are unable to reach a billion requests per second in a commodity server when accessing memory. In particular, we showed that 
state-of-the-art hashtables forfeit core 
functionality, block excessively, and handle 
accesses inefficiently in memory-resident workloads. 
We presented \dlht, a practical non-blocking and memory-aware hashtable. DLHT addresses the issues of existing designs to offer
complete functionality,
fast non-blocking Resizes, 
1.6B requests per second,
and 3.5$\times$ (12$\times$) the throughput of state-of-the-art (open-) closed-address\-ing designs on Gets (Deletes).

\newcommand{\showDOI}[1]{\unskip}
\newcommand{\shownote}[1]{\unskip}

\bibliographystyle{ACM-Reference-Format}
\bibliography{references}



\end{document}